\pdfoutput=1

\documentclass[10pt, conference, compsoc]{IEEEtran}   %IEEE (with arabic numbering)
\usepackage{array}
\usepackage[dvipsnames]{xcolor}
\usepackage{graphicx}
\usepackage{balance}
\usepackage[cmex10]{amsmath}   %IEEE
\usepackage{algpseudocode}
\usepackage[noadjust]{cite}

\usepackage[hyphens]{url}
\usepackage[pdftex,
  pdftitle={Dovetail: Stronger Anonymity in Next-Generation Internet Routing},
  pdfauthor={},
  pdfpagemode=UseOutlines,
  bookmarks,bookmarksopen,
  pdfstartview=FitH,
  colorlinks, linkcolor=black, citecolor=PineGreen, urlcolor=BlueViolet, ]{hyperref}

\newsavebox{\ieeealgbox}
\newenvironment{boxedalgorithmic}
  {\begin{lrbox}{\ieeealgbox}
   \begin{minipage}{\dimexpr\columnwidth-2\fboxsep-2\fboxrule}
   \begin{algorithmic}}
  {\end{algorithmic}
   \end{minipage}
   \end{lrbox}\noindent\fbox{\usebox{\ieeealgbox}}}

\newcommand{\paragraphX}[1]{\vskip 0pt \noindent \textbf{#1} \hskip .05in}

\newcolumntype{C}[1]{>{\centering\let\newline\\\arraybackslash\hspace{0pt}}p{#1}}
\newcolumntype{L}[1]{>{\raggedright\let\newline\\\arraybackslash\hspace{0pt}}p{#1}}
\newcolumntype{R}[1]{>{\raggedleft\let\newline\\\arraybackslash\hspace{0pt}}p{#1}}
\newcolumntype{E}{>{\centering\let\newline\\\arraybackslash\hspace{0pt}}c}

\usepackage[draft,layout={inline}]{fixme}   			% During regular working
\usepackage{soul}          
\renewcommand{\FXLayoutInline}[3]{
	\ifthenelse{\equal{#1}{note}}{\sethlcolor{yellow}}{\sethlcolor{red}}	
	{\bfseries\hl{#2}}
}
\newenvironment{smdescription}
  {\begin{list}{}
     {
	
     \setlength{\topsep}{6pt}
	\setlength{\parsep}{5pt}
     \setlength{\itemsep}{0pt}
     \setlength{\labelwidth}{0pt}
     \setlength{\labelsep}{1ex}
     \setlength{\itemindent}{1ex}
     \setlength{\leftmargin}{0pt}
	}}
  {\end{list}}

\graphicspath{{figures/}}

\begin{document}

\title{Dovetail: Stronger Anonymity in\\Next-Generation Internet Routing}

% IEEE Author block 
%==================
\author{
     \IEEEauthorblockN{Jody Sankey}
     \IEEEauthorblockA{The University of Texas at Arlington\\jody@jsankey.com}
	\and
     \IEEEauthorblockN{Matthew Wright}
     \IEEEauthorblockA{The University of Texas at Arlington\\mwright@cse.uta.edu}
}

\maketitle
\begin{abstract}
Current low-latency anonymity systems use complex overlay networks to
conceal a user's IP address, introducing significant latency and network
efficiency penalties compared to normal Internet usage. Rather than
obfuscating network identity through higher level protocols, we propose
a more direct solution: a routing protocol that allows communication
without exposing network identity, providing a strong foundation for
Internet privacy, while allowing identity to be defined in those higher
level protocols where it adds value.

Given current
research initiatives advocating ``clean slate'' Internet designs, an
opportunity exists to design an internetwork layer routing protocol that
decouples identity from network location and thereby simplifies the
anonymity problem. Recently, Hsiao et al. proposed such a protocol
(LAP), but it does not protect the user against a local eavesdropper or
an untrusted ISP, which will not be acceptable for many users. Thus, we
propose Dovetail, a next-generation Internet routing protocol that
provides anonymity against an active attacker located at any single
point within the network, including the user's ISP. A major design
challenge is to provide this protection without including an
application-layer proxy in data transmission. We address this challenge in path 
construction by using a {\em matchmaker} node (an end host) to overlap 
two path segments at a {\em dovetail} node (a router). The dovetail then 
trims away part of the path so that data transmission bypasses the 
matchmaker. Additional design features
include the choice of many different paths through the network and the
joining of path segments without requiring a trusted third party. We
develop a systematic mechanism to measure the topological anonymity of
our designs, and we demonstrate the privacy and efficiency of our
proposal by simulation, using a model of the complete Internet at the
AS-level.

\end{abstract}

% ACM categories and keywords
%\category {C.2.2}{Network Protocols}{Routing Protocols}
%\terms{Security, Algorithms, Design} (commented to save space)
%\keywords{Anonymity, source routing, next-generation Internet.}

\sloppy
\section{Introduction}\label{sec:intro}

Our society has experienced a dramatic increase in the extent to which daily life is conducted online, with socializing, shopping, learning, and banking via the Internet now an accepted norm. However, parallel advances in technology have enabled widespread tracking, storage, and correlation of our online activities, and business models have evolved for companies to monetize the user data they collect~\cite{mayer2012third,mikians2012detecting}. Taken together, these factors mean that Internet privacy has become a pressing issue, and one that we argue could benefit from technological solutions.

When we use the Internet, a wide range of identifying information is commonly revealed, but one of the hardest forms of identity to remove is that defined by the network routing protocol (\emph{layer~3}), since this identity is used to deliver data. In today's Internet, IP is the primary layer~3 protocol and IP addresses are in every data packet. Recording a user's IP address can allow an adversary to uniquely identify her, link that identity with her online activity, correlate connections to different services, and partially reveal her geographical and network locations. Previous work has proposed \emph{low-latency anonymity systems} to conceal a user's identity~\cite{crowds:tissec,tor-design}, including her IP address. Tor in particular has been adopted by hundreds of thousands of privacy-concious users worldwide~\cite{tor-metrics}. Current anonymity systems, however, work by creating an overlay network on top of the layer~3 protocol, requiring a sequence of IP transmissions to disguise the original sender. This sequential forwarding and the queueing and processing required in intermediary nodes can incur substantial latency and network efficiency penalties.

We prefer an alternative formulation for this problem: Rather than attempting to conceal a global layer~3 identifier by adding complexity in application protocols, we believe that the layer~3 protocol should not reveal a global identity. Instead, we leave identity management to higher layers in the protocol stack, in only those applications where it provides mutual benefit. 

While privacy by itself is unlikely to motivate a change away from IP routing, a range of additional concerns have emerged within the networking field~\cite{paul2011architectures}, including scalability, security, mobility, challenged environments, and network management, leading to major research initiatives investigating ``clean slate'' Internet designs~\cite{find,fire,akari} that could be used to build the {\em next-generation Internet (NGI)}. A wide range of different NGI routing concepts have already been proposed as a result of these activities~\cite{papadopoulos2010greedy,bhattacharjee2006postmodern,pathlet,rfc6830,yang2006source,yang2003nira,zhang2011scion}. Network virtualization research, showcased in testbeds such as GENI~\cite{geni}, offers hope for a progressive transition to a future routing protocol. These initiatives in NGI provide an opportunity to imagine anonymous communications that do not rely on an overlay network.

We thus propose {\em Dovetail}, an NGI routing protocol that prevents
association of source and destination by an attacker located at any
fixed point within the network. Recently, Hsiao et al. proposed LAP, a
lightweight NGI anonymity protocol~\cite{lap}. Unlike LAP, however,
Dovetail provides protection against observation by local eavesdroppers
and by an untrusted ISP, which is a critical requirement for many
privacy-conscious users. 

A major design challenge is to provide this protection without including
an inline application-layer entity (i.e.  a proxy) in data transmission,
which would be much slower than only traversing routers. We address this
challenge in path construction by asking a {\em matchmaker} node (an end
host) to put together two path segments so that they overlap at a {\em
  dovetail} node (a router), and enabling the dovetail to trim away part
of the data forwarding path to remove the matchmaker. This technique is
implemented using public-key operations only at the source and the
matchmaker, while routers use only symmetric encryption and decryption
of short header fields and a simple hash chain. The protocol enables the
choice of many different paths through the network and does not require
a trusted third party.

In brief, our key contributions are: (1) a novel privacy-preserving NGI routing protocol, (2) a systematic mechanism for measuring anonymity in terms of topological identity, and (3) evaluation of our protocol in terms of topological anonymity using an Internet-scale simulation.

The remainder of this paper is structured as follows: Section 2 introduces our objectives and the adversary we design against. Section 3 discusses both source-controlled routing and low-latency anonymity systems, including two systems that we build upon. Section 4 presents the design of Dovetail in a sequence of increasingly detailed perspectives, from the broad design down to the packet structure. Section 5 analyzes the security of our system by considering potential attacks and our defenses, and derives the information available to a passive attacker. Section 6 describes our evaluation of the protocol. Section 7 concludes with a summary of our findings and recommendations for further work.

\section{\label{sec:objectives}Objectives}

In this section, we describe the goals of the system we intend to deliver and the attacker we design against. 

\subsection{\label{sec:anon_objective}Anonymity Objectives}

We refer to the party who initiates a connection as the \emph{source} and the opposite party as the \emph{destination}, although data is able to pass in both directions once the connection is established. Using the terminology of Pfitzmann and Hansen~\cite{anon_terminology}, we aim to provide \emph{unlinkability} between the source and destination, such that no network location is able to sufficiently distinguish whether the source and destination are related, except for the source itself. This implies that network locations with good information on the source identity have little information on the destination identity, and vice versa. Throughout our work, we constrain ourselves to the identifying properties defined at the network layer: network identity and network location, or \emph{topographical anonymity}~\cite{lap}. 

We do not protect the packet contents, which reside in higher network layers and are thus out of scope for this paper. Content should be protected end-to-end using a protocol such as IKEv2, which protects sender and receiver identities~\cite{ikev2}. Such protection is effectively mandatory for strong anonymity protections, as many other forms of Internet identification exist, such as device fingerprinting~\cite{eckersley2010unique} and persistent cookies~\cite{soltani2009flash}. Additionally, higher-level protocols like IKEv2 should be used with restricted options and implementations to limit the possibility of fingerprinting.

\subsection{\label{sec:performance_objective}Performance \& Practicality Objectives}

Any anonymity system must offer acceptable performance in order to gain widespread adoption and thus provide a large set of potential message sources~\cite{econymics}. Performance problems with Tor have been widely discussed, and they are considered an important factor limiting its adoption~\cite{dingledine2009performance,jansenlira}. We aim to provide a lightweight system where all communication for an established connection remains within the core networking infrastructure and occurs at layer~3. This avoids the frequently slow \emph{last mile} connections~\cite{dischinger2007characterizing} in overlay anonymity systems and also the queuing required to move between layers in the protocol stack. Finally, we require that our system provides mechanisms to trade anonymity for performance.

Another key to widespread adoption is recruiting service providers. Our work targets a future Internet, so Dovetail need only compete with other future routing protocols rather than motivate service providers to switch away from IPv4. Today's ISP business models may not apply, but it is unlikely that service providers are willing to spend substantial time and infrastructure for privacy. Our goal is to ensure that costs for service providers are limited, such that benefits for privacy-aware consumers are enough incentive to participate in the protocol. To this end, we recognize that Internet routers have high throughput and low computing resources per flow, so we limit cryptographic operations and avoid maintenance of any per-connection routing state. Additionally, our design does not require significant extra traffic and does not violate basic notions of consumer-provider relationships that exist in today's Internet.

\subsection{\label{sec:attack_model}Attack Model}

Selecting an attack model for anonymity systems is a challenging task in its own right, as the adversary may be different for different users and its capabilities are not known in advance. A few key points guide our decision. First, protecting a low-latency connection from an adversary who can observe traffic at multiple points of the network is very difficult. Tor uses layered encryption and fixed packet sizes to prevent trivial linkability, but this comes with significant expense and does not hide traffic patterns, which are linkable with a small chance of error~\cite{levine2004timing}. Adding sufficient delays and cover traffic to mask traffic patterns is expensive and can be undermined by manipulating the patterns~\cite{rainbow,voip06}. Thus, like both Tor and LAP, we do not aim to protect against these attacks. Second, users may be suspicious of any service provider that can link them with their Internet activities. This applies to anonymity service providers, such as Anonymizer.com, and also to Internet service providers. ISPs have proved to not be fully trustworthy with private browsing data~\cite{sellingClickstream,cableOneSpy}. We therefore aim to prevent any element of our system from being able to deanonymize users. Third, a user's local communication may be subject to eavesdropping, e.g. at a wireless hotspot or by an employer. Unlike LAP, we aim to protect against such adversaries. Fourth, many of the adversaries that we aim to protect against would be capable of various active attacks, such as replay or packet header manipulation, so we also aim to limit the exposure that such attacks might cause.

We thus consider an adversary who is \emph{active} but \emph{local}. Active means the adversary is able to initiate connections and to violate the rules of the protocol for the connections in which she is involved, in addition to passively monitoring these connections. We define local as confined to a single \emph{Autonomous System} (AS) within the Internet. ASes are the level at which routing information and policies are commonly shared, so a compromise in security at one router may affect multiple routers controlled by the same AS. In contrast, in order to span multiple ASes, an attack must either compromise multiple organizations or involve collusion between these organizations. We note that if a particular set of ASes were suspected of collusion, our client logic could easily be modified to include no more than one member of the set in each connection. Our adversary is assumed to have local knowledge of traffic, but global knowledge of the network topology and routing data. 

More concretely, the possible attackers we aim to protect against include: a local eavesdropper, the source ISP, the destination ISP, any single AS in between, any node facilitating our protocol operations, and the destination itself. Thus, we aim for significantly greater protection than LAP or a centralized proxy server like Anonymizer.com. 

Given that we only protect against a single observation point, we offer no protection against attacks that require multiple observation points, even though such attacks may be practical for state-level adversaries~\cite{entropist} or Internet exchange points~\cite{murdoch-pet2007}. In common with LAP, but not Tor, we do not try to prevent trivial linkability based on packet contents and sizes. This means that linking attacks with multiple observation points need lower computational and storage resources and succeed with fewer observations than against Tor. Additionally, if both the source and destination are customers of the same ISP, it is simple for the ISP to correlate traffic. Again, Tor provides basic protection that makes this attack slightly harder, while both LAP and Dovetail provide no protection.

Finally, given recent revelations about the NSA and GCHQ, some will
argue that an anonymity system should protect against such adversaries,
who may view traffic over undersea cables~\cite{mti,undersea} and perhaps
could target Internet Exchanges (IXs)~\cite{murdoch-pet2007}. We believe
that Dovetail is flexible enough to accommodate such consideration into
the routing mechanisms, but we leave the design and evaluation of such
an extension to future work.

\section{Background}\label{sec:background}

In this section, we cover two research areas of direct relevance to our problem: source-controlled routing protocols and low-latency anonymity systems. Within each area, we describe a proposal that our design builds upon.

\subsection{Source-Controlled Routing}\label{sec:source_routing}

One theme spanning a number of next-generation Internet routing proposals is that of source-controlled routing, in which the originator of a data packet has some control over the route it takes, usually using routing control information carried in the data packet. In some protocols, the source has influence over the route but not complete control~\cite{yang2006source,zhang2011scion}; in others, the source explicitly declares the route that should be taken~\cite{pathlet,yang2003nira}. As we explain in Section~\ref{sec:design_space}, this ability to express a route at the source has benefits for anonymity in addition to the robustness and flexibility considerations that initially motivated the research. IPv4 includes source control options~\cite{rfc791}, but these can be used to violate firewall rules and routing policies and are normally disabled. We assert that these security concerns stem from inadequate consideration of security in the design of IPv4~\cite{bellovin1989security} and need not apply to all implementations of source-controlled routing.

\subsubsection{Pathlet Routing}\label{sec:pathlets}

Pathlet routing~\cite{pathlet} is one example of a source-controlled routing system. Each entity within a network defines a number of virtual nodes (or \emph{vnodes}) and advertises path segments (or \emph{pathlets}) that pass between these vnodes. Vnodes are a virtual construct, so a single physical router may process packets for multiple vnodes, or a single vnode may be distributed across multiple physical routers. Each vnode is defined by a forwarding table containing the set of allowed outgoing pathlets. All packets arriving from a particular communication peer are processed by one vnode whose forwarding table defines the set of allowed routes for that peer. The pathlet protocol provides an expressive system that is able to represent many different types of routing policy, although more complex routing policies require a greater number of vnodes. Godfrey et al. demonstrate that \emph{local transit} policies (i.e. policies only dependent on ingress and egress points for their own network) may be represented efficiently and independently of the total network size~\cite{pathlet}. This is in contrast to the BGP exterior gateway protocol~\cite{rfc4271} commonly used today, where the forwarding information base must scale linearly with the total number of advertised IP prefixes.

To send a packet, the source assembles a list of adjacent pathlets defining the intended route and includes this list in the packet header. Each pathlet is represented by a variable length Forwarding ID (\emph{FID}), an index into the forwarding table of the vnode that defined the pathlet. When a vnode receives a packet, it removes the first FID and uses this as an index into its forwarding table to determine which link the packet should be sent over. Only legal routes are defined in the forwarding tables. Therefore, unlike in BGP, it is impossible to violate the routing policy by invoking unannounced routes, since no such routes exist. When a vnode learns of adjacent single-hop pathlets, it may choose to aggregate these together into a composite pathlet. The translation from this composite into single-hop pathlets is stored in the forwarding table, so when a packet requests the composite path, the forwarding table is used to restore the component pathlets. Pathlet routing moves the responsibility for network route creation from the network infrastructure to the end hosts originating traffic. This means the large routing information base embodying network topology need only be consulted each time a new route is constructed, and not each time a packet is forwarded. It also provides flexibility for an end host to control how its packets will traverse the network. A source may rapidly select alternative routes to achieve better performance or compensate for network failures, instead of waiting minutes for an exterior gateway protocol to converge upon a new route.

\subsection{\label{sec:low_latency_anon}Low-Latency Anonymity Systems}

Previous works have introduced a variety of low-latency anonymity systems whose response time is sufficient for general-purpose interactive use, including Web browsing. A wide range of practical systems have been proposed and some of these have been fielded~\cite{boyan1997anonymizer,crowds:tissec,freedom2-arch,tor-design}. Current low-latency anonymity systems may be categorized as either centralized or distributed. Centralized systems pass all traffic though an anonymizing proxy, which must be trusted. Distributed systems overlay an additional network on top of the current layer~3 protocol and therefore require multiple IP transmissions to deliver each packet from source to destination. These multiple transmissions, together with processing inside the intermediate hosts, contribute to latencies that are substantially higher than Internet usage without anonymization~\cite{PanchenkoPR08}.

\subsubsection{\label{sec:lap}Lightweight Anonymity and Privacy}

In Lightweight Anonymity and Privacy (LAP)~\cite{lap}, Hsiao et al. propose the anonymity scheme that inspires our work. Their protocol relies upon \emph{packet-carried forwarding state}, where the information required to deliver a packet is stored within the packet itself. To establish a connection, the source constructs a packet containing a sequence of \emph{autonomous domains} (ADs) describing the route. As each AD receives the packet, it encrypts its own routing instruction using a private symmetric key and forwards the packet to the next AD. Once a connection has been constructed in this manner, data may be exchanged between the endpoints using the resulting encrypted header. Each path construction request contains a nonce that influences the encryption process, allowing a source to construct multiple unlinkable connections over the same route by using different nonces. Header padding may be included to partially obfuscate the path length. During construction, each AD on the path learns the identity of all ADs that follow it but not the identity of the ADs before it. Some information on predecessor identity may be inferred based on knowledge of the preceding AD, network topology, routing policy, observed header length, and observed response time, but these are not quantified. LAP assumes the user's own ISP is trustworthy, and it provides no protection of source-destination unlinkability against a local eavesdropper or an observer at the source ISP. Given previous well-publicized ISP indiscretions~\cite{sellingClickstream,cableOneSpy} and the possibility of a hacker infiltrating this single point of failure, it seems unlikely that privacy-conscious users will share this assumption. 

Other than LAP, $\mathsf{AND\bar{a}NA}$ is the only other
next-generation Internet anonymity protocol that we know
of~\cite{andana}. It is only designed for named-data networks and it is
built using onion routing, both of which are very different from
Dovetail.

\section{Design}\label{sec:design}

In this section, we first provide context for our design point and then describe the protocol from four different perspectives in increasing detail.

\subsection{Layer 3 Anonymity Design Space}\label{sec:design_space}

To provide a broadly applicable anonymity system, we assert that any layer~3 solution should provide two features:

\paragraphX{Deviation from shortest path.} An eavesdropper can measure information on the length of the network path before and after her\footnote{Throughout this paper, we use the genders of the standard security actors: The source, Alice, is female, the destination, Bob, is male, and the attacker, Eve, is female.} vantage point. If a routing protocol always selects the shortest possible route, then when the shortest route between participants is significantly shorter or longer than the Internet average, the protocol will reveal this abnormal distance and limit their anonymity. 

\paragraphX{Partitioned routing information.} When the routing information is stored as a single field, such as an IP address, any entity with access to the field may calculate the destination identity. When routing information is divided across multiple fields, then an entity must access multiple fields to learn the destination identity. Fields may be protected independently to prevent this access.

Source-controlled routing is useful since it accommodates both of these features: when the source of a message can dictate a path, she is free to pick one that is not the shortest, and she may express the path as a separate instruction for each entity along the route. Dovetail uses construction requests that are traceable in the forwards direction as presented by Hsiao et al.~\cite{lap}. Our detailed design builds upon the pathlet routing protocol presented by Godfrey et al.~\cite{pathlet}. Pathlet routing works well for our system, but we are not reliant on any unique feature of this protocol. The principles we describe could be applied to any protocol that provides complete control over the selected route and a wide range of allowable routes.

\subsection{Network Model}\label{sec:network_design}

We propose a clear distinction in routing at the AS boundary; each AS should expose the minimum number of vnodes and pathlets necessary to satisfy its routing policies. This distinction provides two practical benefits: First, minimizing the number of externally visible vnodes reduces the size of the routing information base that must be held in end hosts. Second, distinguishing between internal and external connectivity allows an AS to retain a flexible and dynamic internal routing policy. As with BGP, adjacent ASes share routing information to establish the network topology. This communication should be secured against MITM attacks that could selectively filter the topology. We assume that hosts know the numeric identity of the vnodes they wish to contact. An equivalent to DNS would be required to translate human-readable identities into vnode identities. The translation service itself could be accessible using Dovetail to protect privacy, but is outside the scope of our current work.

\begin{figure}[htb]
\centerline{\includegraphics[width=3.4in]{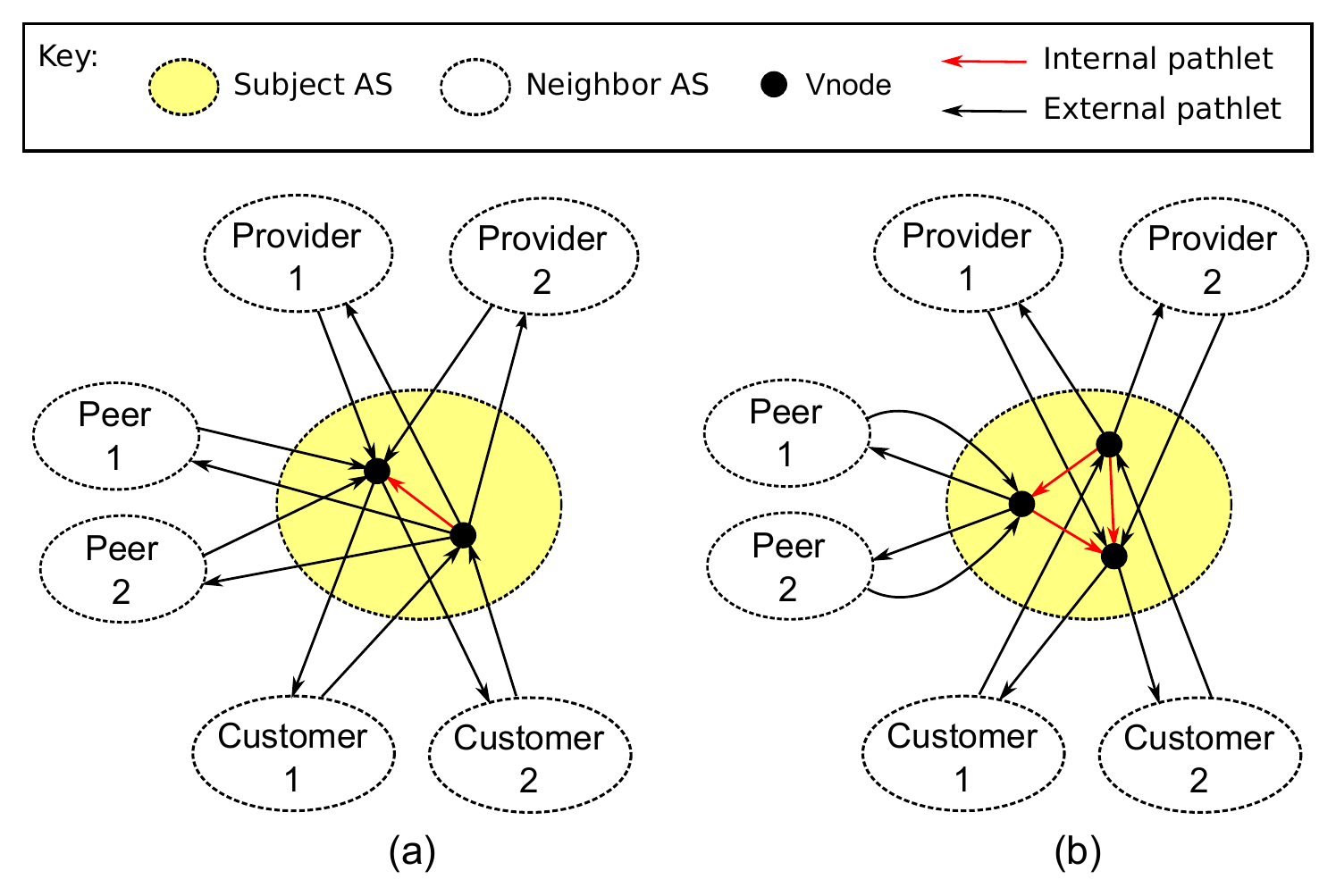}\vspace{-2.0ex}}
\caption{AS vnode and pathlet structure by routing policy: a) Strict valley-free, b) Loose valley-free}\label{fig:vnodes}
\end{figure}

The most common form of routing policy used in the Internet today is \emph{valley-free routing}~\cite{gao2001inferring}, which reflects the contractual relationships between ASes. A \emph{customer} AS is one who pays a \emph{provider} AS to forward its traffic, while two ASes with a \emph{peer} relationship will each forward each other's traffic without payment. In valley-free routing, each AS will only forward traffic when there is a financial incentive to do so, i.e. when the traffic originates from or is destined for a paying customer. As illustrated in Figure~\ref{fig:vnodes}a, two vnodes are required per AS to enforce this strict definition of a valley-free routing policy: one to receive traffic from customer ASes and one to receive traffic from peer and provider ASes. Although valley-free routing is common, BGP allows for arbitrarily complex routing policies and valley-free routing is not ubiquitous~\cite{giotsas2012valley}. In particular, there are a growing number of Internet exchange points (IXPs), which offer ASes the ability to peer with each other and thereby save money~\cite{ixp}. Most transit and access provider ASes will peer with any non-customer AS~\cite{lodhi2014open}. This suggests that peering is compatible with ASes' incentives and is likely to continue to be common.

We thus consider a slightly relaxed routing policy, which we refer to as \emph{loose valley-free}. In this scheme, an AS will allow traffic to pass between its peers. The AS would not receive payment from a customer for performing this service, but also is not required to make a payment and could avoid payments at other times if peers provide a reciprocal service. As shown in Figure~\ref{fig:vnodes}b, three vnodes are required per AS to enforce a loose valley-free routing policy: one to receive traffic from customer ASes, one to receive traffic from provider ASes, and the third to send and receive peer traffic.

For good anonymity properties as described in
Sec.~\ref{sec:outpath_design}, Dovetail relies on a modest fraction of
ASes to use the loose valley-free policy or other policies that are less
strict than valley-free routing. If all ASes use strict valley-free
routing, Dovetail still provides anonymity, but with smaller anonymity
sets.

\subsection{Path Construction}\label{sec:dovetail_design}

\begin{figure}[tbh]
\centerline{\includegraphics[width=3.35in]{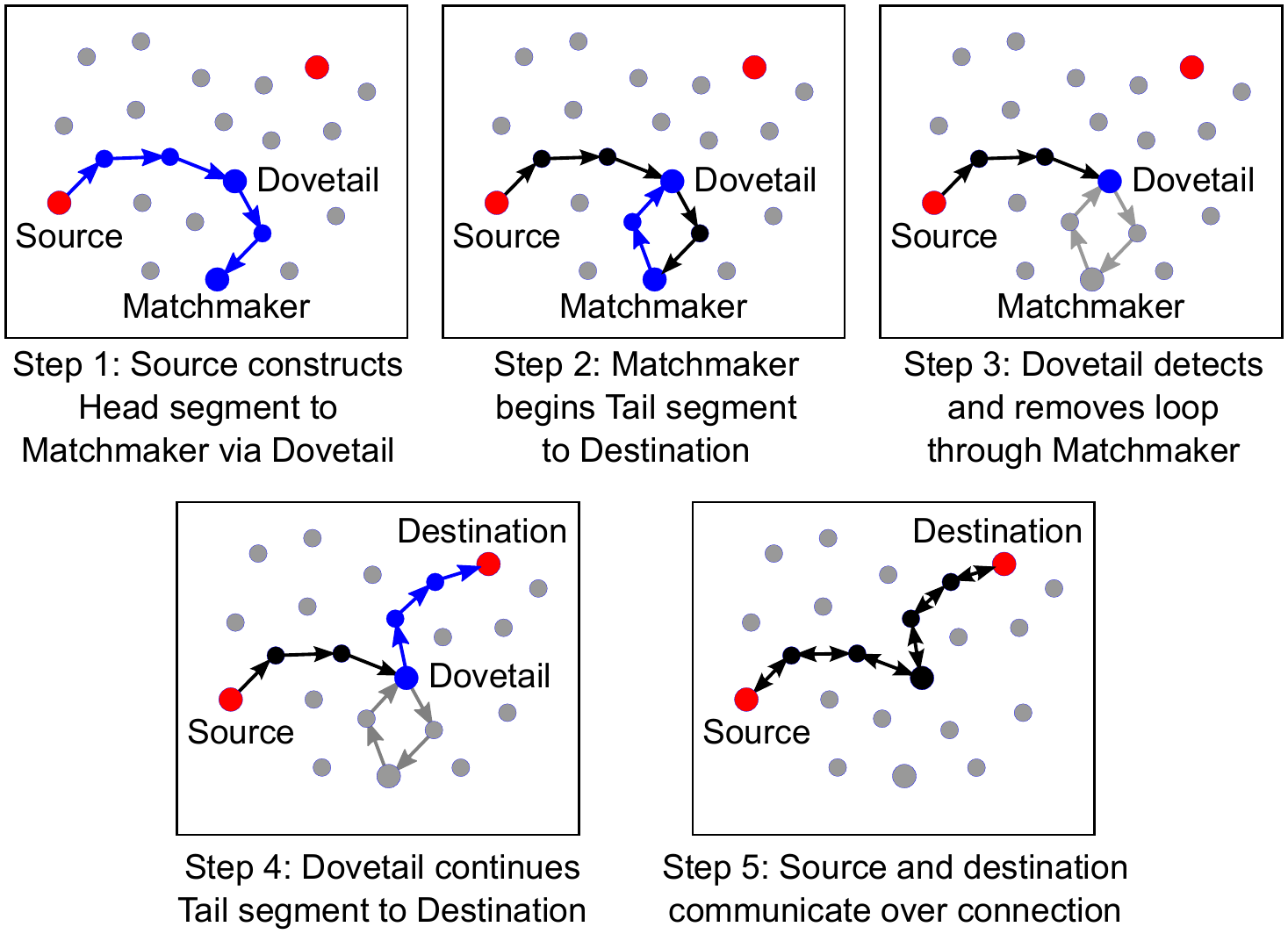}\vspace{-1.0ex}}
\caption{Construction of a Dovetail connection}\label{fig:nopseudonym_construction}
\end{figure}

A Dovetail path comprises multiple \emph{path segments}. As we explain in Section~\ref{sec:outpath_anon}, an AS that is present on a path segment may learn the identity of subsequent ASes and its direct predecessor, but not earlier ASes. Figure~\ref{fig:nopseudonym_construction} illustrates the Dovetail path creation process. 

The path cannot be constructed directly from the source to the destination, since the source's ISP would be able to link source and destination. Instead, we make use of a randomly selected, untrusted third-party vnode called the \emph{matchmaker}. This matchmaker may either be an end host or functionality exposed by a service provider. Providing matchmaker services should cost little relative to enabling our protocol in routers. The identities of vnodes willing to act as matchmakers could be distributed as a part of routing information maintenance. 

The source encrypts the identity of the final destination using a public key for the matchmaker and builds a \emph{head} path segment to the matchmaker, who then extends the path to the destination with a \emph{tail} path segment. Here, the source ISP no longer learns the identity of the destination, only of the matchmaker. The matchmaker learns the identity of the destination, but cannot identify the source through the intervening ASes. The source may learn the matchmaker's public key without compromising anonymity by requesting a signed certificate over the same path used to establish the connection. To improve performance and minimize the trust we must place in the matchmaker, we prefer that the matchmaker not be involved in the exchange of data. Therefore, we require that the head and tail segments cross at some vnode, referred to as the \emph{dovetail}\footnote{We use the term to reflect a dovetail joint in carpentry, where two elements are joined securely and compactly}. The source encrypts the identity of the dovetail and provides it to the matchmaker for inclusion on the tail segment. The dovetail detects the crossing condition and joins the two segments, removing the loop in the path along with the matchmaker. 

The tail path segment would ideally be selected by the source, but the source does not have complete knowledge of distant Internet topology. The matchmaker has sufficient knowledge to construct a path to the destination, but the user's anonymity can be degraded if an AS appears on both the head and tail segments, and therefore we prefer that the tail segment avoids ASes already used on the head segment. Providing a list of head ASes to the matchmaker would reveal substantial information on the source identity, so instead we ask the matchmaker to return a set of potential tail routes that the source selects from. The source then sends its choice to the matchmaker to complete the route. For brevity, we do not include this tail selection mechanism in our description of the packet design, but we do examine its effect on anonymity in Section~\ref{sec:dovetail_eval}.

\subsection{Segment Route Selection}\label{sec:outpath_design}

A source-controlled routing system may attempt to obfuscate path length, but an attacker located on the path will be able to infer some information about her distance to the source and destination through round trip timing, packet length and structure analysis, and active probing. We prefer a system that is robust even when an attacker learns path length to one that relies on keeping it hidden. For the remainder of the discussion, we assume the attacker has perfect knowledge of the number of ASes preceding and following her own, but limit the value of this knowledge through a non-deterministic path selection process. 

Our mechanism for routing each path segment is based upon the principle of \emph{path diversity}, where a large number of possible paths may be taken from any given source to any given destination. We note that this is beneficial for the robustness of the system in addition to its anonymity. To achieve path diversity, each host must have a comprehensive, but not necessarily complete, map of the network. We extend the pathlet routing protocol by exporting extra pathlets in addition to the shortest path tree (SPT). The optimal set of additional pathlets depends on network size and topology, but our experiments show that for the current Internet, is it appropriate to export 50\% of the SPT size, selecting pathlets closest to the sender. An important consequence is that routing knowledge varies across the network, and so any assessment of available path options can only be made in the context of the vnode (in our case, the source or the matchmaker) selecting the path. Maintenance of routing information in response to network changes could be performed using path vector distribution methods similar to BGP~\cite{rfc4271}, but this is not relevant to the anonymity properties of the system and so is not discussed further. 

When a host constructs a path segment, it will normally have a wide range of options available with different \emph{costs}, where we define cost as the number of times the route changes AS. Other cost metrics such as latency or bandwidth could also be integrated into the protocol. The distribution of options across cost reflects the network topology between the source and destination. Selecting a random path uniformly from among the complete set of options would reveal information about this distribution, such as picking the most common path cost most frequently, and thus leak information about the topology. Instead, we use a {\em cost window approach}: we select a path by first selecting a path cost and then randomly selecting one of the paths at this cost. We explain this scheme further in Section~\ref{sec:outpath_eval}.

\subsection{Data Packet Structure}\label{sec:packet_design}

Dovetail extends the basic packet format used in pathlet routing, providing a set of different packet types and processing algorithms for each type. These algorithms provide the following security properties:

\begin{enumerate}
\item An AS does not learn the identity of ASes before its immediate predecessor.
\item AS routing information is protected by a key known only to the AS.
\item Different connections travelling over the same route do not produce the same ciphertext.
\item The final ciphertext for each AS depends on the entire path.
\item An AS may only create a removable loop in the path when given access to privileged information. This information is only given to the matchmaker.
\end{enumerate}

\begin{figure}[htb]
\centerline{\includegraphics[width=3.15in]{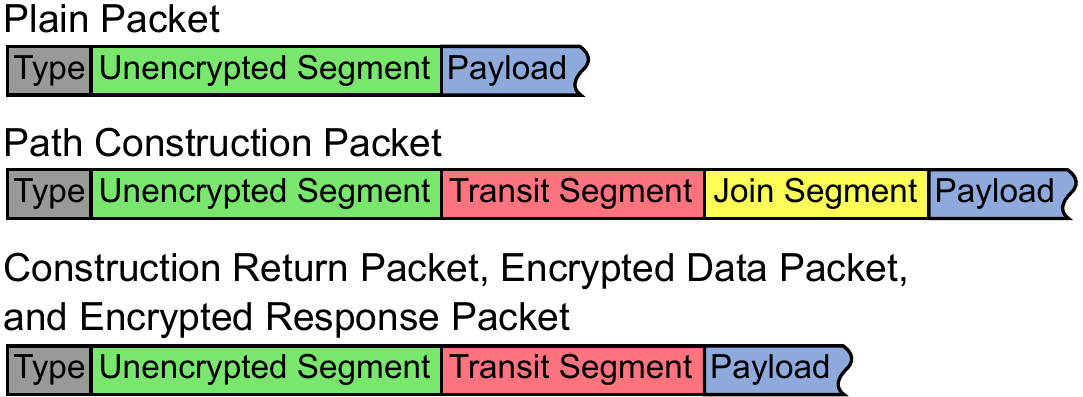}}
\caption{Structure for each packet type}\label{fig:packet_structure}
\end{figure}

\begin{figure}[htb]
\centerline{\includegraphics[width=3.4in]{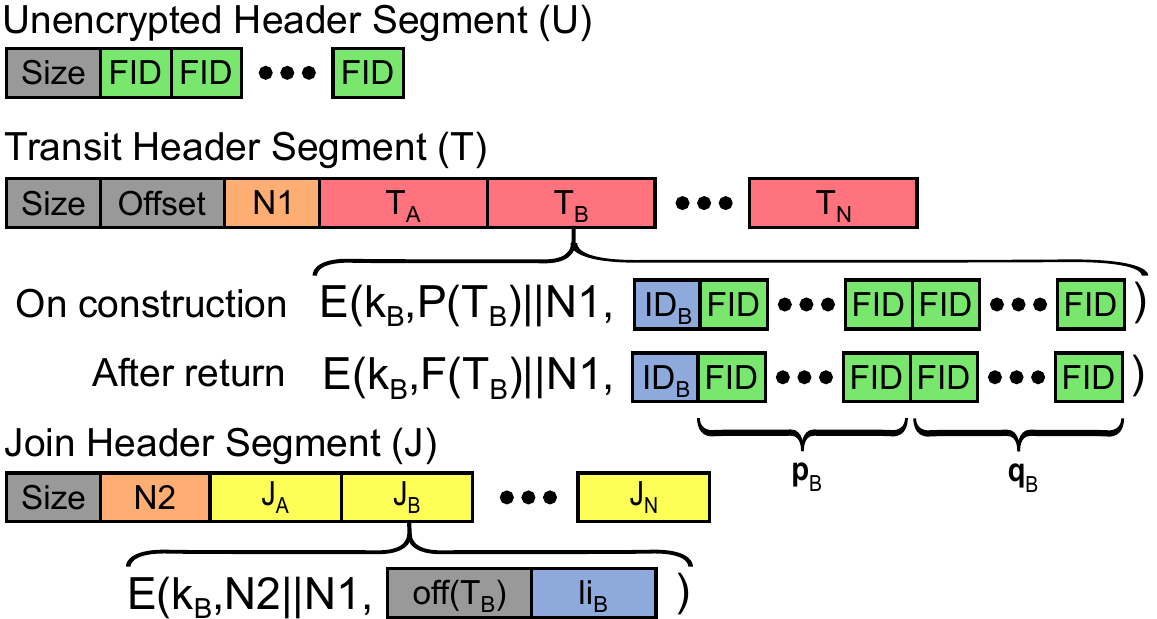}}
\caption{Structure for each header segment type}\label{fig:segment_structure}
\end{figure}

Each Dovetail packet contains a type identifier, followed by one or more variable-length header segments, followed by the payload. Table~\ref{tbl:packet_notation} summarizes the notation we use in this section. Figure~\ref{fig:packet_structure} presents the packet types in terms of their header segments, while Figure~\ref{fig:segment_structure} presents the structure of each header segment. 

\begin{table}[b]
\normalsize
\caption{\label{tbl:packet_notation}Packet structure notation}
\begin{center}
\begin{tabular}{|c l|}\hline

\bfseries Term & \bfseries Definition \\ \hline\hline
$U$ & Unencrypted packet segment. Stores \\
    & partial path during construction and \\
    & while traversing an AS. \\
$T$ & Transit packet segment. Stores complete\\
    & bi-directional path in encrypted form. \\
$J$ & Join packet segment. Facilitates routing \\
    & loop detection during path construction.\\
\hline
$T_A$ & Transit entry for AS $A$.\\
$J_A$ & Join entry for AS $A$. \\
$\mathit{id}_A$ & Identifier for AS $A$. \\
$\mathit{li}_A$ & Transmission link used to enter AS $A$. \\
$k_A$ & Symmetric encryption key for AS $A$. \\
$\boldsymbol{p}_A$ 	& Sequence of pathlets traversing AS $A$ \\
				& in the forward direction. \\
$\boldsymbol{q}_A$ 	& Sequence of pathlets traversing AS $A$ \\
				& in the reverse direction. \\
$\mathit{m}_A$ & Maximum number of bits required to \\
               & represent any path traversing AS $A$. \\
\hline
$\mathit{N1}$, $\mathit{N2}$ & Nonce values. Initialized randomly then \\
                             & modified during path construction. \\
\hline
off($x$)   & Offset of field $x$ from segment start. \\
P($T_A$)   & Transit entries preceding $T_A$. \\
F($T_A$)   & Transit entries following $T_A$. \\
$H(x)$ & Cryptographically secure hash of $x$. \\
$E(k,v,x)$ & Encryption of $x$ with key $k$ and IV $v$. \\
$D(k,v,x)$ & Decryption of $x$ with key $k$ and IV $v$. \\
\hline
\end{tabular}
\end{center}
\end{table}

Dovetail does not perform cryptographic operations on the packet payload, only on routing information in the packet header. Restricting our operations to a small number of bytes in the header allows for fast operation within the routing infrastructure. Privacy-preserving transport layer protocols, such as IPsec, may be used to encrypt or authenticate data when needed. To create a Dovetail connection, the source issues a path construction packet, leading the destination to respond with a construction return packet. Once the response is received, the source and destination may continue to communicate over the connection using a sequence of encrypted data and encrypted response packets, each containing the transit segment created during path construction. Vnodes need not store any per-connection state in order to exchange data. We now discuss each of these packet types in turn, in order of increasing complexity. 

\begin{figure}[b]
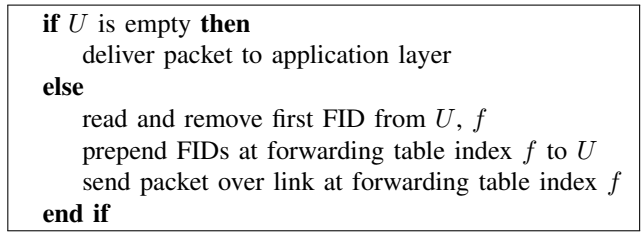

\begin{boxedalgorithmic}
  \If{$U$ is empty}
    \State deliver packet to application layer
  \Else
    \State read and remove first FID from $U$, $f$\;
    \State prepend FIDs at forwarding table index $f$ to $U$\;
    \State send packet over link at forwarding table index $f$\;
  \EndIf
\end{boxedalgorithmic}
\caption{\normalsize\label{fig:plain_algorithm}Plain Packet processing}
\end{figure}

\begin{figure}[thb]
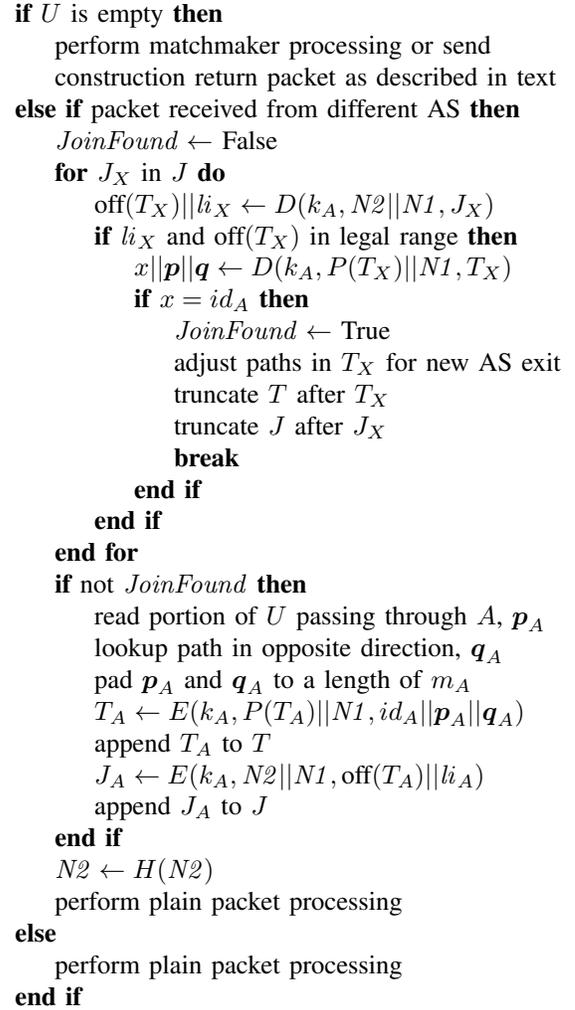

\begin{boxedalgorithmic}
\If{$U$ is empty}
	\State perform matchmaker processing or send 
	\State construction return packet as described in text
\ElsIf{packet received from different AS}
	\State $\mathit{JoinFound} \leftarrow$ False
	\For{$J_X$ in $J$}
		\State $\mbox{off}(T_X)||\mathit{li}_X \leftarrow D(k_A,\mathit{N2}||\mathit{N1},J_X)$
		\If{$\mathit{li}_X$ \upshape and $\mbox{off}(T_X)$ in legal range}
			\State $x||\boldsymbol{p}||\boldsymbol{q} \leftarrow D(k_A,P(T_X)||\mathit{N1},T_X)$
			\If{$x = id_A$}
				\State $\mathit{JoinFound} \leftarrow$ True
				\State adjust paths in $T_X$ for new AS exit
				\State truncate $T$ after $T_X$
				\State truncate $J$ after $J_X$
				\State \textbf{break}
			\EndIf
		\EndIf
	\EndFor
	\If{not $\mathit{JoinFound}$}
		\State read portion of $U$ passing through $A$, $\boldsymbol{p}_A$
		\State lookup path in opposite direction, $\boldsymbol{q}_A$
		\State pad $\boldsymbol{p}_A$ and $\boldsymbol{q}_A$ to a length of $m_A$
		\State $T_A \leftarrow E(k_A,P(T_A)||\mathit{N1},id_A||\boldsymbol{p}_A||\boldsymbol{q}_A)$
		\State append $T_A$ to $T$
		\State $J_A \leftarrow E(k_A,\mathit{N2}||\mathit{N1},\mbox{off}(T_A)||\mathit{li}_A)$
		\State append $J_A$ to $J$
	\EndIf
	\State $\mathit{N2 }\leftarrow H(\mathit{N2})$
	\State perform plain packet processing
\Else
	\State perform plain packet processing
\EndIf
\end{boxedalgorithmic}
\caption{\normalsize\label{fig:construction_algorithm}Path Construction processing within AS $A$}
\end{figure}

\subsubsection{Plain Packet}
A plain packet follows the underlying pathlet routing protocol without the anonymity features introduced by Dovetail. A plain packet only contains an unencrypted header segment defining a string of forwarding table indexes (i.e., FIDs). Upon receipt of a Plain Packet, each vnode follows the standard pathlet packet processing algorithm (Figure~\ref{fig:plain_algorithm}).

\subsubsection{Path Construction Packet}
A path construction packet is used to establish an encrypted path through a matchmaker and dovetail. The source initially constructs the packet containing a path to the matchmaker in the unencrypted segment but no entries in the transit and join segments. Upon receipt of a path construction packet, the receiving vnode follows the algorithm shown in Figure~\ref{fig:construction_algorithm}. Each time the packet enters a new AS, tests for a loop join condition by searching for previous join entries encrypted with a matching key and nonces. If the join condition is met, the AS deletes all transit and join entries after its first inclusion on the path. If the join condition is not met, the AS adds a new encrypted transit entry to describe the shortest path through the AS in both directions, and a new fixed-length join entry that enables location of the variable-length transit entry. Both the join and transit entries are encrypted with a symmetric key known only to the AS. This key is used for all connections through the AS but should be changed periodically to limit the impact of compromise. A secure hash function is applied to $\mathit{N2}$ at each new AS, similar to a Lamport hash chain~\cite{lamport1981password}, preventing an attacker from predicting previous nonce values and therefore from creating artificial joins. 

When a path construction packet is received with an empty unencrypted segment, the response depends on the packet payload. If the payload is empty, which is the case at the destination, the receiver replies with a construction return packet. This packet is created from the path construction packet by discarding the join segment, adding a payload containing $\mathit{N1}$, and then updating $\mathit{N1}$ to $H(\mathit{N1}||T_A...T_N)$. If the payload contains a \emph{continuation request}, which happens at the matchmaker, the matchmaker decrypts the continuation request with its public key to learn the identities of the destination and dovetail, the cost to dovetail, and a prior hash of $\mathit{N2}$. The matchmaker uses this information to compute possible tail paths and return these to the source for selection as discussed in Section~\ref{sec:outpath_design}. The matchmaker then removes the packet payload, creates a new unencrypted header segment containing the selected tail path, and sets $\mathit{N2}$ to the prior hash received from the source. This last step ensures that the dovetail receives the same value of $\mathit{N2}$ on both the head and tail segments.

\subsubsection{Construction Return Packet}
A construction return packet is used to return the selected path to the source. The transit segment is re-encrypted during return using the algorithm in Figure~\ref{fig:construction_return_algorithm}, such that the final ciphertext depends upon the entire selected path. 

\begin{figure}[thb]
\begin{boxedalgorithmic}
\If{packet received from different AS}
	\State use $\mathit{offset}$ to locate $T_A$ within $T$ 
	\State $x||\boldsymbol{p}_A||\boldsymbol{q}_A \leftarrow D(k_A,P(T_A)||\mathit{payload},T_A)$
	\If{$x \neq \mathit{id}_A$}
		\State \textbf{return}\Comment{Invalid packet}
	\EndIf
	\State replace current entries in $U$ with $\boldsymbol{q}_A$
	\State $T_A \leftarrow E(k_A,F(T_A)||\mathit{N1},\mathit{id}_A||\boldsymbol{p}_A||\boldsymbol{q}_A$
	\State $\mathit{offset} \leftarrow \mathit{offset} - (\mbox{len}(\mathit{id}_A) + 2m_A)$
\EndIf
\State perform plain packet processing
\end{boxedalgorithmic}
\caption{\normalsize\label{fig:construction_return_algorithm}Construction Return processing within AS $A$}
\end{figure}

\begin{figure}[tb]
\begin{boxedalgorithmic}
\If{packet received from different AS}
	\State use $\mathit{offset}$ to locate $T_A$ within $T$ 
	\State $x||\boldsymbol{p}_A||\boldsymbol{q}_A \leftarrow D(k_A,F(T_A)||\mathit{N1},T_A)$
	\If{$x \neq id_A$}
		\State \textbf{return}\Comment{Invalid packet}
	\EndIf
	\State replace current entries in $U$ with $\boldsymbol{p}_A$
	\State $\mathit{offset} \leftarrow \mathit{offset} + (\mbox{len}(\mathit{id}_A) + 2m_A)$
\EndIf
\State perform plain packet processing
\end{boxedalgorithmic}
\caption{\normalsize\label{fig:data_algorithm}Encrypted Data processing within AS $A$}
\end{figure}

\subsubsection{Encrypted Data Packet}
An encrypted data packet is used by the source to send data over an existing Dovetail connection. The packet is constructed with an empty unencrypted segment and the previously received transit segment, with an offset set to zero. Each time an encrypted data packet enters a new AS, the current transit entry is decrypted and the unencrypted segment is replaced with the forward path from this entry, as described by the algorithm in Figure~\ref{fig:data_algorithm}.

\subsubsection{Encrypted Response Packet}
An encrypted response packet is used by the destination to send data over an existing Dovetail connection. The packet is constructed with an empty unencrypted segment and the previously received transit segment. The offset is set to the end of the last transit entry. Each time an encrypted response packet enters a new AS, the current transit entry is decrypted and the unencrypted segment is replaced with the reverse path from this entry, as described by the algorithm in Figure~\ref{fig:response_algorithm}.

\begin{figure}[tbh]
\begin{boxedalgorithmic}
\If{packet received from different AS}
	\State use $\mathit{offset}$ to locate $T_A$ within $T$ 
	\State $x||\boldsymbol{p}_A||\boldsymbol{q}_A \leftarrow D(k_A,F(T_A)||\mathit{N1},T_A)$
	\If{$x \neq id_A$}
		\State \textbf{return}\Comment{Invalid packet}
	\EndIf
	\State replace current entries in $U$ with $\boldsymbol{q}_A$
	\State $\mathit{offset} \leftarrow \mathit{offset} - (\mbox{len}(\mathit{id}_A) + 2m_A)$
\EndIf
\State perform plain packet processing
\end{boxedalgorithmic}
\caption{\normalsize\label{fig:response_algorithm}Encrypted Response processing within AS $A$} 
\end{figure}

\section{Security Analysis}\label{sec:analysis}

In this section, we assess the security of the Dovetail protocol. We
consider a range of anonymity attacks that might be applied against the
protocol and then analyze the information available to a passive
attacker at each point in the network. We end with brief discussions of
timing attacks and attacks on availability and integrity.

\subsection{Attacks on Anonymity}\label{sec:anonymity_attack}
As Dovetail is lightweight, it does not protect against attacks that
succeed against an overlay system like Tor. In particular, an entity who
can observe traffic at multiple points in the connection can link both
of those points, which can link a source to her destinations. In
Dovetail, this is trivial, as the packet contents are not encrypted
differently at different points in the network. In Tor, however, timing
analysis can enable this linking with high
accuracy~\cite{voip06,rainbow}. Other attacks that rely on multiple
points of observation, such as selective denial of service~\cite{sdos}
and predecessor~\cite{wals08} attacks will be just as effective in
Dovetail. Additionally, Dovetail is vulnerable to the same types of
side-channel attacks that impact
Tor~\cite{ajaxchannels,throughput,latency,torta05,congestion-longpaths}.

Beyond this, however, we need to examine additional attacks that could
leverage the unique aspects of the Dovetail protocol. The primary
information available to a passive attacker in the network is the cost
to the source and destination and the preceding and following ASes in the
path, and we examine the affect of these on anonymity in Sections 
\ref{sec:outpath_anon} and \ref{sec:dovetail_anon}. Timing attacks are
considered briefly in Section~\ref{sec:timing_anon}. Other attacks
include:

\begin{smdescription}

\item[Observe or correlate packet content.] Dovetail is a layer~3
  protocol and does not provide any protections for the data it
  carries. In cases where packet content would reveal identity, or where
  confidentiality is important, a higher layer protocol such as IKEv2
  should be used to provide encryption~\cite{ikev2}.

\item[Correlate connections from a source.] Each connection includes a
  source-defined nonce, $\mathit{N1}$. When the source changes this
  nonce, a different ciphertext will be produced, preventing an observer
  from associating multiple connections over the same path from their
  header content. When connections between a source-destination pair are
  distinctive, and may hence be correlated by some other property, the
  source could reuse the same matchmaker and path to prevent
  intersection and predecessor attacks.

\item[Replay packets.] A replayed packet will take the same path as its
  original transmission and therefore not provide an attacker with new
  information. An adversary might try to probe for the source by
  prepending an unencrypted path to a recorded packet, but each AS
  empties the unencrypted segment on receipt to prevent this attack.

\item[Probe for a later AS.] To determine the destination of an observed
  connection, an attacker may try to construct many new connections
  through the same dovetail and search for matches in the header
  ciphertext. Dovetail protects against this attack by including a hash
  of the entire path in the IV for encrypted transit segments. Any
  change in the selected path will therefore perturb the ciphertext for
  all segments.

\item[Probe for an earlier AS.] The joining of a Dovetail path provides
  confirmation that the joining AS appeared on the path twice, and an
  attacker may wish use this feature to probe for suspected
  predecessors. During connection construction, an attacker may attempt
  to extend the path to a suspect and then back to herself, where she
  could observe whether a join occurred. Our use of hash chaining on
  $\mathit{N2}$ prevents this attack, since the attacker cannot
  replicate the nonce initially presented to the suspect. The matchmaker
  is provided with an earlier nonce to create a legal join and may
  perform some probing, but this is heavily constrained by the
  dovetail-matchmaker cost limit.

\item[Matchmaker intersection.] The matchmaker provides the source with
  a set of possible tail segments from which the source picks one. Since
  the source will not select an AS already on the head segment,
  including it's own ISP, the matchmaker could try to offer tail
  segments that help it isolate possible source ASes. In particular, if
  there is a source AS of interest $A$, then the matchmaker could pick
  tail segments that include likely ASes between itself and $A$. If the
  source avoids these tail segments, it adds to the likelihood that the
  source is in $A$. However, fully unmasking the source AS with this
  type of intersection attack would require a large number of
  requests. As matchmakers are selected randomly from a large set, an
  attacker located at any particular matchmaker is unlikely to receive
  many connection requests from the same source.

\item[Modify the requested path.] An AS along the path could modify the
  unencrypted header segment to alter the route taken for the remainder
  of the path segment, but gains little from doing so. All vnodes along
  a path segment can identify the destination, and earlier vnodes have a
  better knowledge of the source. Thus, an attacker that places herself later
  in the same path segment does not learn any additional information regarding
  source or destination. 

\item[Modify the tail path.] The matchmaker could use a different
  tail option than that selected by the source. However, the matchmaker
  does not learn whether unselected paths were acceptable and cannot
  identify the source and so cannot predict whether a particular path
  will be bad for that source. A matchmaker could speculatively route
  all connections through a particular ISP to allow identification of
  any sources within that ISP. This attack may be effective given a
  sufficient number of matchmakers, but widespread collusion falls 
  outside our attack model.

\end{smdescription}

\subsection{Single Segment Anonymity Analysis}\label{sec:outpath_anon}

We now examine the source and destination anonymity at each point along
a single path segment as a step towards analyzing the complete path. We
present two approaches for assessing single segment topological
anonymity: a simple and efficient approach based on set size and a more
accurate approach based on entropy.

\subsubsection{Anonymity Set Size}\label{sec:outpath_setsize_anon}

\begin{sloppypar}
Consider an attacking AS, $AS_i$, located at cost $i$ in a path segment
$AS_0,AS_1, \cdots AS_{i-1},AS_i \cdots AS_n$. This AS can identify its
predecessor, $AS_{i-1}$, but cannot directly identify earlier ASes,
since their routing instructions have been encrypted. The remaining
portion of the segment is not encrypted, and therefore all following ASes
and the destination are known. $AS_i$ can accurately measure the 
path cost from the source to itself by using the length of the join
segment in the packet header, and thus it can deduce the total path
cost $n$. We use $S_y^{AS_x}$ to represent the set of possible sources
that have a shortest path to $AS_x$ of cost $y$.
\end{sloppypar}

\begin{smdescription}

\item[Shortest Path.] When the shortest path is used, an adversary
knows that the observed cost for $AS_{i-1}$ and for all subsequent ASes
must be the shortest cost from the true source. The set of possible
sources is therefore the intersection of the possible source sets for
each observed AS:
\begin{equation*}
	\mbox{sources}(\mathit{shortest\,path},AS_i) = \bigcap\limits_{j=(i-1)}^n S_j^{AS_j} 
\end{equation*}
If $n$ is close to the minimum or maximum cost present in the Internet,
then few sources will fall into this intersection, leading to an
uncomfortably small anonymity set.

\item[Cost Window Selection.]  As an alternative, we propose a
\emph{cost window} selection algorithm to select uniformly at random a
length between some global minimum $p$ (or the shortest path cost if it
is greater) and some global maximum $q$. Ideally for anonymity, $q$
should be greater than the maximum shortest path cost in the
network. Given cost window selection, an attacker cannot make any
statement about the possible message senders, except that each must be
able to form a path of the observed length to the observed predecessor
$AS_{i-1}$. Our experiments show that as long as at least a small
fraction of ASes use a loose valley-free routing policy, long path
choices are plentiful. This means that, in most cases, a source can
produce a path at any given cost greater than the minimum. We find this
to be true 96\% of the time when 10\% of ASes are loose
valley-free. Making an approximation that this is always true, then the
set of possible sources is simply the union of the sources at every
distance less than or equal to the observed value:
\begin{equation*}
	\mbox{sources}(cost\,window,AS_i) = \bigcup\limits_{j=0}^{i-1} S_j^{AS_{i-1}} 
\end{equation*}
By examining the relationship to $A_{i-1}^{AS_{i-1}}$, it is clear that:
\begin{equation*}
\mbox{sources}(\mathit{shortest\,path},AS_i) \subseteq \mbox{sources}(cost\,window,AS_i)
\end{equation*}
Hence, cost window selection provides an equal-sized or larger source
anonymity set in all cases. In addition, $n$ is capped to a minimum of
$q$, and therefore very short path costs with unavoidably small
anonymity sets will never be generated. Using cost window selection,
even sources with a low shortest path cost will occasionally select very
long paths, and therefore the average latency will be higher than in
shortest path selection. A non-uniform probability distribution can be
used to control how frequently larger costs are selected and limit this
performance penalty.

\end{smdescription}

\begin{table}[htb]
\caption{\label{tbl:entropy_notation}Effective anonymity set size notation}
\begin{center}
\begin{tabular}{|c L{2.5in}|}\hline

\bfseries Term & \bfseries Definition \\ \hline\hline
  $S$ & The set of all possible sending vnodes\\
  $s$ & The true source vnode\\
  $d$ & The destination vnode\\
  $\rho$ & The path selected to send data from $s$ to $d$\\
  $a$ & An attacking vnode located on $\rho$ and wishing to identify $s$ within $S$\\ 
  $a'$ & The vnode on $\rho$ immediately preceding $a$\\
  $\mbox{t}_\rho(x)$ & The portion of path $\rho$ after vnode $x$\\
  $\lambda_\rho(x)$ & The cost along $\rho$ from the source to $x$\\
  $\mbox{OBS}_a$ & The set of observations available to $a$\\
  $R(x,y)$ & The set of paths from $x$ to $y$\\
  $R(x,y,\lambda)$ & The set of paths from $x$ to $y$ of cost $\lambda$\\
\hline
\end{tabular}
\end{center}
\end{table}

\subsubsection{Effective Anonymity Set Size}\label{sec:outpath_entropy_anon}

The preceding analysis is simple and efficient for an attacker to
compute using only shortest path distances, but with complete knowledge
of routing tables, she can achieve a better result. The probability that
a given potential source selected the observed path depends on the
available path options for that source, and thus it is not uniform
across the set of potential senders. These differences in probability
allow calculation of an effective anonymity set size based on the
entropy of the distribution. The notation we use in this section is
summarized in Table~\ref{tbl:entropy_notation}. For each possible source
$t$, the probability of selecting the observed path cost
$P_{cost\:match}(t)$ depends only on the path selection algorithm and
the presence or absence of paths in $R(t,d)$ at each cost, not the
number or definition of these paths nor the location of the attacker. Section~\ref{sec:outpath_eval} presents a series of different
options for path length selection, along with their cost selection
probabilities. Once a cost has been selected, a path of this cost is
chosen randomly from the available set. The probability that $t$ chose a
path matching the observations is thus given by the fraction of paths
that both place the observed predecessor at the observed cost and that
match the observable portion of the path:
\begin{equation*}
\begin{split}
P_\mathit{predecessor\:match}(t) = \\ 
   \frac{|\sigma, \sigma \in R(t,d,\lambda_\rho(d)) \wedge \lambda_\sigma(a') = 
\lambda_\rho(a') \wedge \mbox{t}_\sigma(a') = \mbox{t}_\rho(a')|} {|R(t,d,\lambda_\rho(d))|}
\end{split}
\end{equation*}
We assume that the a priori probability $P_{apr}(t=s)$ of each vnode $t$
being the source of a given message is uniform. The probability that $t$
is the true source given an attacker's observation may then be
calculated by Bayes theorem:
\begin{equation*}
P(\mbox{OBS}_a | t=s) = P_\mathit{cost\:match}(t) \times P_\mathit{predecessor\:match}(t) 
\end{equation*}
\begin{equation*}\label{eqn:bayes}
P(t=s | \mbox{OBS}_a) = \frac{P(\mbox{OBS}_a | t=s) \times P_{apr}(t=s)}
			{\sum\limits_{i \in S} P(\mbox{OBS}_a | i=s) \times P_{apr}(i=s)} 
\end{equation*}
Finally, we may use this set of potential source probabilities to
compute an effective source anonymity set size based on the entropy of
the distribution by using the technique proposed by Serjantov and
Danezis~\cite{serjantov2003entropy}:
\begin{equation*}
S = -\sum_{t \in S}P(t=s | \mbox{OBS}_a)\;log_2(P(t=s | \mbox{OBS}_a))
\end{equation*}

\subsection{\label{sec:dovetail_anon}Complete Path Anonymity Analysis}

A passive adversary who observes the construction of a Dovetail path
segment has full knowledge of the remainder of the segment and partial
knowledge of the segment source. She may learn further information from
observing the return path. We now discuss the complete set of
information available regarding source and destination identity at each
location on a Dovetail path. When a measurable cost is available, a set
of possible identities can be built using the techniques defined above
in Section~\ref{sec:outpath_anon}.

\begin{smdescription}

\item[Source Identity.] The source identity is known to the source
ISP. An attacker at each subsequent AS towards the matchmaker (which
includes the dovetail node) can use its knowledge of the preceding AS
identity, cost from the source, and all subsequent pathlets up to the
matchmaker to limit the possible source identities. At the matchmaker
itself, for paths of more than three or four hops, the number of
possible sources should be quite large. After the matchmaker, the amount
of information about the source will be even less.

\item[Destination Identity.] The destination identity is known to
every ISP from the matchmaker to the destination ISP due to the
construction request. Any AS on the head segment out to the
matchmaker, but that does not appear on the data path, has no knowledge
of the destination. Between the source and the dovetail, an attacker can
measure the cost from the destination to her own AS using the data
return path. If the attacker is able to guess which AS on the head
segment serves as the dovetail, she can infer cost from the destination
to the dovetail. 

As intended, locations where the source is easily identified have little
information about the destination and vice versa. The dovetail is the
closest AS to the source that learns destination identity; it is
typically the strongest location for a passive attacker. To avoid
elevating the capability of an attacker located at the dovetail AS, we
require that this AS only appear on the head segment once. Any other
AS that appears twice in a given segment gains no additional information
from its second inclusion. 

Each segment of the dovetail path serves a purpose in maintaining a
particular anonymity property; this should be considered when setting
the segment length. The head segment must be long enough to conceal
source identity from the dovetail, and the tail segment must be long
enough to conceal destination identity from the source ISP. Finally, we
note that uniform random selection of the matchmaker, uncorrelated with
either the source or destination, is effective in isolating the
anonymity properties of our system. An AS on the head segment can
identify the matchmaker, but this does not help to identify the
destination; an AS on the tail segment may be able to identify the
matchmaker, but this does not help to identify the source.

\end{smdescription}

\subsection{\label{sec:timing_anon}Response Timing Attacks}

The path diversity used to select each segment should hinder an attacker's
ability to identify participants from response timing data. Each potential
source could have used one of many thousand possible routes to reach the
destination, and each of these routes has its own latency distribution. The
superposition of these distributions blurs the range of possible response times
for a source significantly when compared to shortest path routing and thus
makes distinguishing between different sources harder.

A performance-optimized version of Dovetail could consider geographical 
distance or latency in its selection of a matchmaker, which in turn would 
change the anonymity properties of the protocol. 

We consider this integration of performance and anonymity concerns in response 
to network latency information to be a rich avenue for further study.

\subsection{\label{sec:availability_attack}Availability and Integrity Attacks}

\begin{smdescription}

\item[Violate routing policy.] As with pathlets, all forwarding tables
  entries are valid expressions of the routing policy, and hence it is
  not possible to construct a path that violates this policy.

\item[Construct arbitrarily long paths.] Our packet design constrains
  the maximum length of both encrypted and unencrypted packet header
  segments and thus limits the longest path an adversary intending to
  waste resources can construct.

\item[Overload a matchmaker.] A matchmaker could be overloaded by
  sending a large number of continuation requests, but matchmakers are
  distributed throughout the network and the effect on clients is minor
  if the first matchmaker they contact is unavailable.

\item[Overload a routing vnode.] Our forwarding operations are simple and
  intended to operate at the full data rate of a router. Connection 
  construction requires more operations, but a maximum connection rate
  could be enforced to constrain this resource utilization. 

\item[Pollute routing tables.] Securing the integrity of routing table
  updates is an important requirement, but we do not consider the
  routing maintenance protocol here hence it is outside the scope of our
  work.

\item[Modify packet contents.] Dovetail is a layer~3 protocol and does
  not provide any protections for the data it is used to carry. In cases
  where integrity is important, a higher layer protocol should be used
  to provide authentication.

\item[Discard packet data.] If the quality of service provided by a
  connection drops below some threshold, this would be observed as a
  failure, for which the recommended remedy is to reconnect over a
  different path. Paths are constructed by random selection from the
  available routes, and so this reconnection is likely to remove any
  intermediate AS discarding data.
\end{smdescription}

\section{Evaluation}\label{sec:evaluation}

Our proposal is evaluated primarily by simulation, using a model of the complete Internet at the AS level. In this section, we first introduce our simulation and input data, then discuss the anonymity and cost results for path segments and for complete paths, and conclude by estimating a variety of resource requirements for our system.

\subsection{Simulation Scope}\label{sec:sim_scope}

\begin{figure*}[t]
\centerline{\includegraphics[width=6.5in]{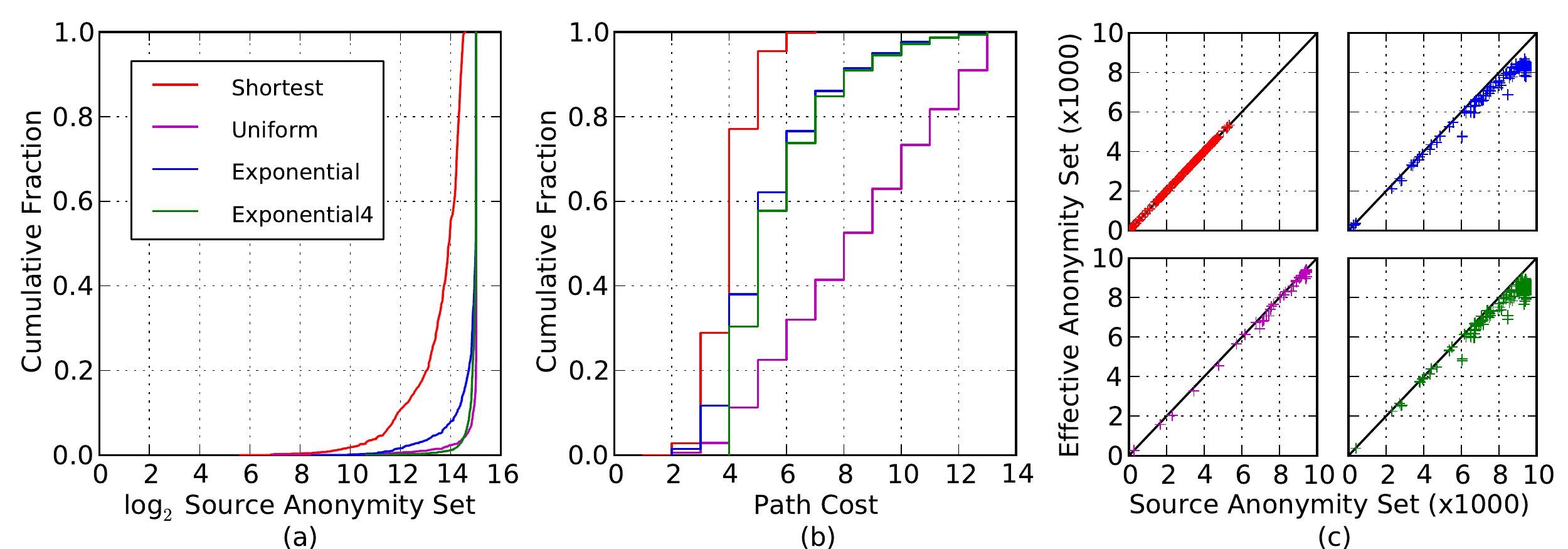}}\vspace{-1.0ex}
\caption{\label{fig:outpath_mondo_comparison}Comparison of segment path selection algorithms: a) Cumulative source anonymity set size distribution, b) Cumulative cost distribution, c) Source anonymity set size vs. effective set size}\vskip -8 pt
\end{figure*}

Our simulation models a network of ASes, each containing up to three routing vnodes plus host vnodes to represent its end users and matchmaking capability. ASes are connected by pathlets that codify their contractual arrangement; customer, provider, or peer. All pathlets within an AS have a cost of zero and all pathlets between different ASes have a cost of one. We simulate the exchange of routing information at initialization, leading to a unique routing perspective for each AS that contains all routing vnodes but not all pathlets. Separately, we simulate packets at a bit level during a connection, allowing us to test header design to ensure that routers and the matchmaker could correctly run the protocol.

Our Internet topology is derived from the CAIDA {\em inferred AS relationship dataset}~\cite{caida}. The dataset contains \emph{sibling} relations, which permitted infinitely long valley-free routes in some circumstances. To avoid optimistic bias, we replaced all sibling relationships with the more restrictive \emph{peer} relationship. This reclassification causes 5.5\% of the network to lose complete reachability, so we disallow traffic originating from or terminating at these ASes. We consider each AS without customer ASes to be a service provider for end users and add a host vnode to represent these users. Ideally, we would model the number of users, but accurate ISP customer size data are not available. Rather than risk skewing our conclusions, we restrict ourselves to measuring anonymity based on the number of possible source or destination ISPs, recognizing that some ISPs are far larger than others. 

We consider a mixture of ASes following the strict and loose valley-free routing policies defined in Section~\ref{sec:network_design}. Experimentation shows that when all ASes follow strict valley-free, the number of routing options is limited, but introducing even a small proportion of loose valley-free ASes leads to far greater diversity. 10\% loose valley-free ASes gives a median of 91,000 options for each path, and we use this topology for the remainder of our evaluation. Studies show that strict valley-free routing is not universal today~\cite{giotsas2012valley}, but we acknowledge that our selection of 10\% is arbitrary.

\subsection{Single Segment Performance}\label{sec:outpath_eval}

\begin{figure*}[htb]
\centerline{\includegraphics[width=6.8in]{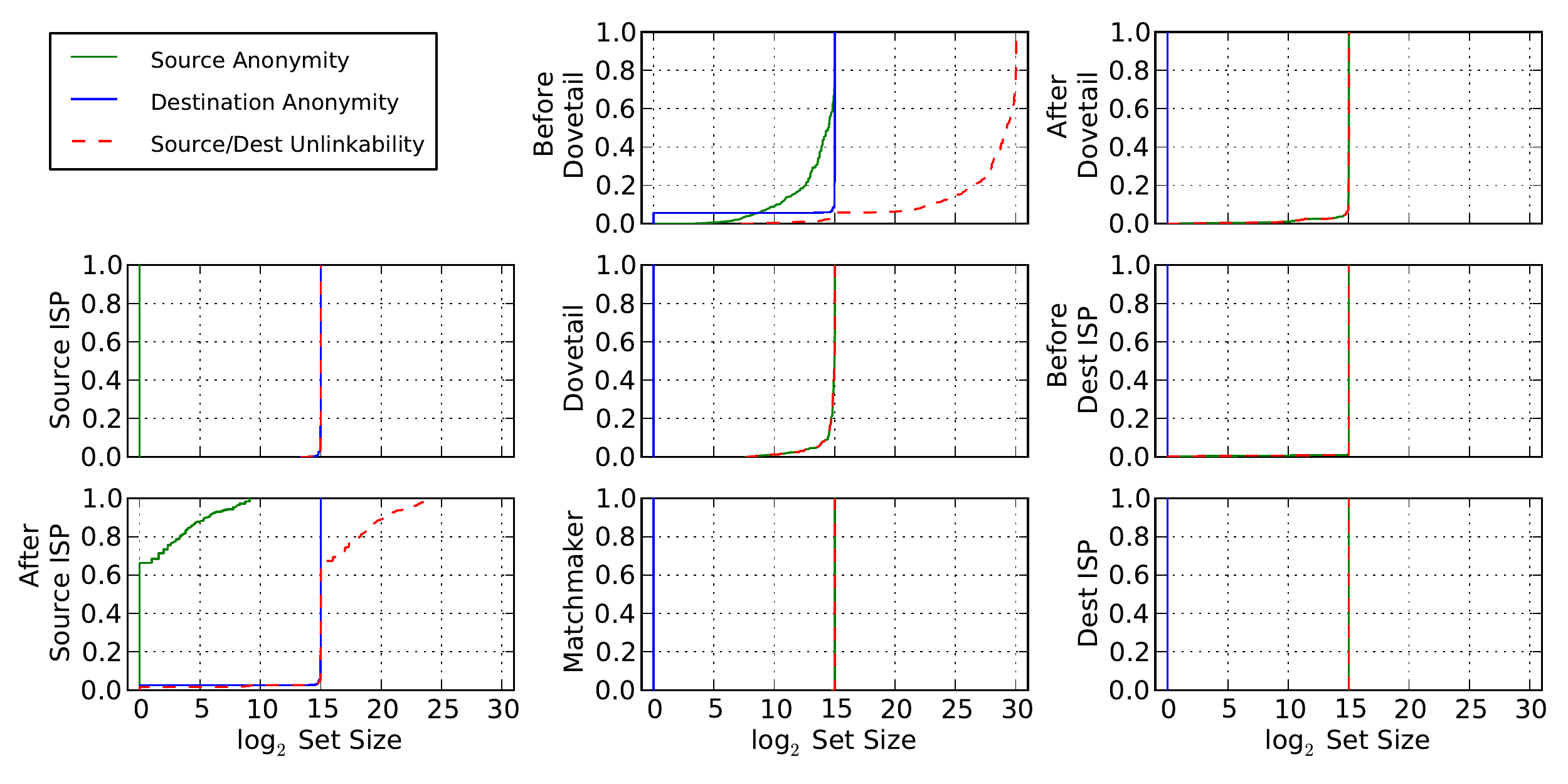}\vspace{-2.0ex}}
\caption{\label{fig:dovetail_notail}Source and destination anonymity set size along the complete path}\vspace{-5 pt}
\end{figure*}

To select a path segment, the source compiles a set of available routes using a modified depth first search. Our implementation limits this set to a maximum cost of 13, based on the longest distance present in the network, and also a maximum of 20,000 routes at each path cost to limit computation. We first select a cost from the set of \emph{available costs} (i.e. costs with at least one route) and then select a random route of this cost. We present results for the following four path selection algorithms, expressed in terms of the probability $P(\lambda)$ that each will select a particular available cost $\lambda$. In all cases, $k$ is a constant specific to the algorithm:

\begin{smdescription}
\item[Shortest.]Shortest possible path is selected in all cases: $P(\lambda) =$ 1 if $\lambda = \lambda_{shortest}$, or 0 otherwise.
\item[Uniform.]Each cost is equally likely: $P(\lambda) \propto 1$.
\item[Exponential.]Longer paths are less likely: $P(\lambda) \propto k^\lambda$.
\item[Exponential4.]Longer paths are less likely, but costs under four are prohibited: $P(\lambda) = 0$ if $\lambda<4$, or $P(\lambda) \propto k^\lambda$ otherwise.
\end{smdescription}

Figure~\ref{fig:outpath_mondo_comparison}a presents a distribution of the cost produced by these algorithms for random source-destination pairs, while Figure~\ref{fig:outpath_mondo_comparison}b presents the source anonymity set size distribution as measured by an attacker at the destination. These results demonstrate that all the algorithms that produce a non-deterministic cost succeed in achieving a meaningful anonymity improvement over the shortest path algorithm, by three bits of entropy in the best case and over one bit for the majority of cases. Algorithms that select long paths more frequently achieve better anonymity but result in a higher average cost. The change from Exponential to Exponential4 is particularly striking, showing a dramatic improvement in worst case anonymity from the exclusion of short paths, with only a modest cost increase. The Exponential4 algorithm results in an average cost approximately 25\% greater than shortest path routing, and yet it achieves an anonymity set containing over half the network in 98\% of the tests. 

In Figure~\ref{fig:outpath_mondo_comparison}c, we compare the set size metric with the \emph{effective} set size metric. This data is from a smaller network, approximately one quarter of the full Internet, and is presented on a linear scale. The data shows that our non-deterministic methods have a slightly smaller effective set size once their cost selection probability is fully incorporated, but this reduction is small. The largest reduction in set size is around 0.35 bits. This shows that cost window selection is effective at limiting information gain by the attacker.

\subsection{Complete Path Performance}\label{sec:dovetail_eval}

We now evaluate the anonymity and cost properties of complete
paths. Dovetail includes parameters that users can configure to trade
performance against anonymity. Our objective here is to demonstrate the
anonymity limit of this sliding scale, but many users will prefer a
lower setting. The parameter settings we use are:
\begin{smdescription}
\item[Dovetail to Matchmaker Cost = Two.] Provides strong limits on matchmaker capability without requiring that dovetail and matchmaker are adjacent.
\item[Source to Matchmaker Algorithm = Exponential6.] Effectively delivers Exponential4 at the dovetail.
\item[Dovetail to Destination Algorithm = Exponential4.] Shown in
  Section~\ref{sec:outpath_eval} to provide near maximum anonymity.
\end{smdescription}

\begin{figure}[h]
\centerline{\includegraphics[width=3.2in]{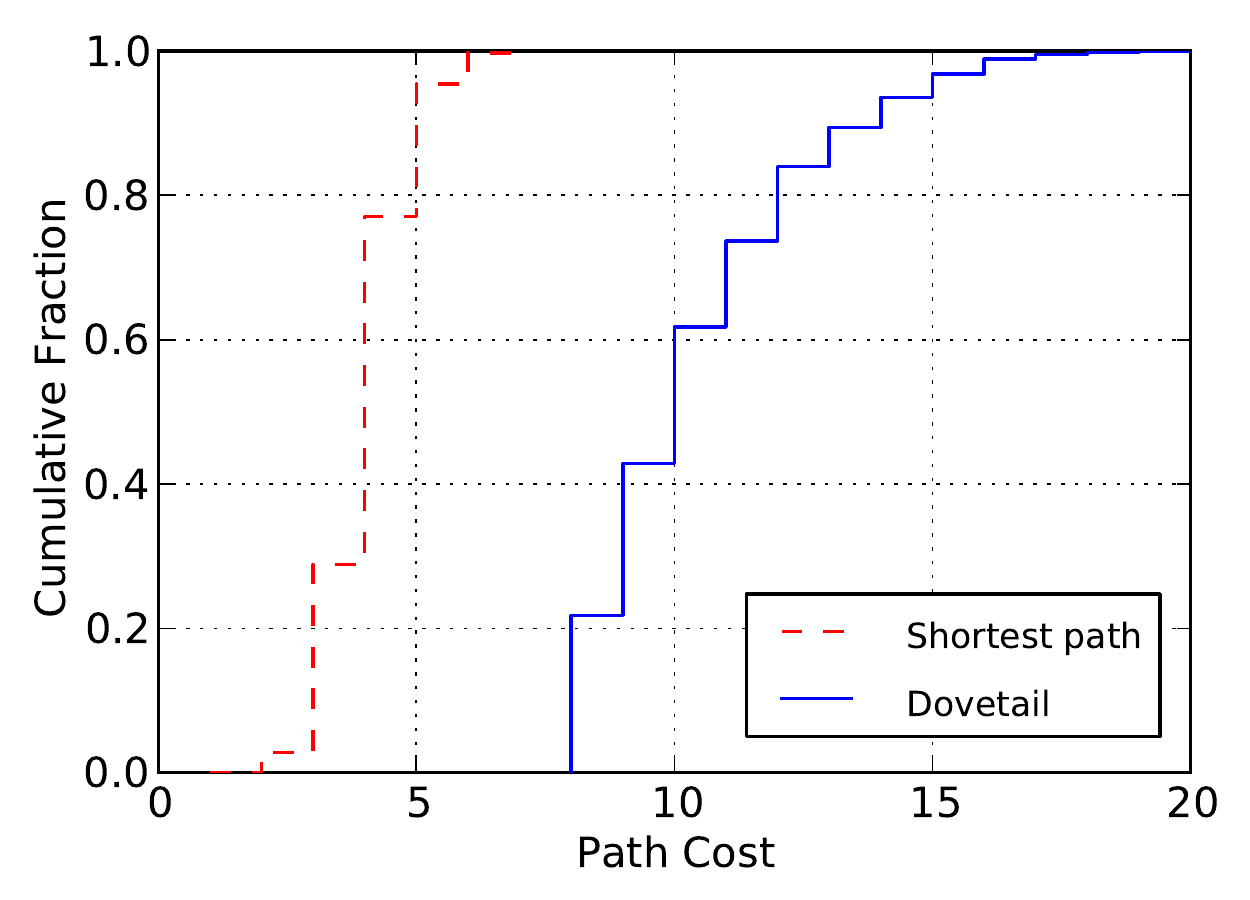}\vspace{-1.5ex}}
\caption{\label{fig:cost_comparison}Cost distribution for complete path}\vskip -12 pt
\end{figure}

In our experiment, we select source and destination hosts at random and
construct a dovetail path between them. The matchmaker generates 
eight tail path options and the source selects one from this set. 
Where possible, the source selects an option that does not reuse a head 
AS, but in 23\% of paths constructed all options required such 
reuse.\footnote{We plan in future work to develop a heuristic to select
  dovetail vnodes with a lower probability of reuse.} We measure the
source and destination anonymity set size observable by an attacker at
each location in the path. Random selection of a matchmaker decouples
the source and destination anonymity sets, and therefore we can also
consider the \emph{source-destination unlinkability}, i.e. the number of
potential source-destination pairs associated with an observed
connection, to be the product of the source and destination anonymity
set sizes. Figure~\ref{fig:dovetail_notail} presents the distribution
of these three properties at a series of key locations along the path,
and Figure~\ref{fig:cost_comparison} presents the cost distribution,
with the cost of shortest path routing included for comparison with
IP and LAP.

Each successive step adds ambiguity to the source identity. At the dovetail AS,
source anonymity is approximately equal to network size in 80\% of cases.
Destination identity is known at the dovetail and all subsequent locations, but
locations prior to the dovetail are unable to calculate a meaningful
destination identity. No location except the source is able to clearly link
source and destination. The AS immediately preceding the dovetail is most
likely to be duplicated in head and tail segments, being adjacent to an AS
that is always present in both. As illustrated by the destination anonymity for
``Before Dovetail'', this occurred in 5\% of our experiments. The dovetail may
partially calculate source identity in around 20\% of cases, but this is
limited to around one thousand possible source ISPs, each containing many
users. Figure~\ref{fig:cost_comparison} shows that a Dovetail path passes
through approximately 2.5 times more ASes than the shortest path routing
used in the current Internet. This is a modest penalty when compared to
the prevailing option for anonymity today; an anonymous circuit in Tor
typically passes through three relays for a total of four IP paths,
including six more last-mile connections than a direct path, and incurs
additional processing and queuing delays at each relay.

\subsection{Resource Utilization}\label{sec:resource_util}

Rather than proposing a near-term solution, we aim to show that privacy
is a feasible feature to include in future routing protocol
designs. Nevertheless, we now briefly consider a variety of resource
requirements to demonstrate that implementation would be feasible.

\begin{smdescription}
\item[Host memory utilization.] Each Dovetail host must maintain a model
  of the Internet to generate routes. In the 2012 dataset we use there
  are 252,666 visible pathlets, of which an average of 22\% are known,
  requiring 680kB.
\item[Router memory utilization.] A Dovetail forwarding table scales
  with the number of local peers and not the total number of Internet
  prefixes as with BGP. All forwarding information is carried by the
  packet itself, and so a router need not store any information per
  connection. 
\item[Router latency.] The only cryptographic operation required to
  forward a data packet is a symmetric decryption of one word. This is
  the same task performed by LAP; Hsiao et al. measure an additional
  latency of under one microsecond in a software-based implementation of
  their system~\cite{lap}.
\item[Transmission efficiency.] A Dovetail packet must specify a
  complete path rather than only an endpoint, potentially leading to
  large headers and low efficiency. The average header length in our
  experiments is 92 bytes. Given an MTU of 1500 bytes, this represents
  a 3.5\% reduction in payload compared to IPv6. LAP would require 
  a 60 byte header.
\end{smdescription}

\balance
\section{Conclusion}\label{sec:conclusion}

In this paper we presented Dovetail, a next-generation Internet routing protocol, and have demonstrated that it provides a workable solution for anonymity at the network layer. The overhead is approximately 2.5 times that of shortest path routing when configured to provide near complete anonymity against our chosen attacker, and we include mechanisms to exchange anonymity for performance. We have demonstrated key aspects of the feasibility and effectiveness of this direction and hope this this motivates serious consideration of privacy as a requirement in the development of other next-generation routing protocols. 

We do not advocate an Internet without identity. Rather, we propose that identity exposure be reserved for substantial relationships and avoided for ephemeral relationships. This research direction is still in its infancy, and much work remains. We believe the most pressing matter is to develop diversity in the solution space by considering how privacy-preserving features might be integrated into other leading layer~3 proposals.  Beyond this, there are user interface questions to explore in the provision of clear and meaningful network privacy choices and in characterizing the user response to these choices. This characterization will provide a better understanding of the resource requirements for network layer anonymity systems and of the motivations for service providers.  Finally, further work remains with the Dovetail protocol. Network latency information should be incorporated into both the path selection algorithms and the anonymity assessments. Our evaluation should be extended to a wider range of network assumptions and additional privacy/performance settings.

% IEEE in the compsoc conference mode this was necessary to prevent item spacing
\newcommand{\BIBdecl}{\setlength{\itemsep}{0.25 em}}

\bibliographystyle{IEEEtran}   %IEEE
%\bibliographystyle{abbrv}      %ACM
%\balance
\bibliography{../dovetail}

\begin{thebibliography}{10}
\providecommand{\url}[1]{#1}
\csname url@samestyle\endcsname
\providecommand{\newblock}{\relax}
\providecommand{\bibinfo}[2]{#2}
\providecommand{\BIBentrySTDinterwordspacing}{\spaceskip=0pt\relax}
\providecommand{\BIBentryALTinterwordstretchfactor}{4}
\providecommand{\BIBentryALTinterwordspacing}{\spaceskip=\fontdimen2\font plus
\BIBentryALTinterwordstretchfactor\fontdimen3\font minus
  \fontdimen4\font\relax}
\providecommand{\BIBforeignlanguage}[2]{{%
\expandafter\ifx\csname l@#1\endcsname\relax
\typeout{** WARNING: IEEEtran.bst: No hyphenation pattern has been}%
\typeout{** loaded for the language `#1'. Using the pattern for}%
\typeout{** the default language instead.}%
\else
\language=\csname l@#1\endcsname
\fi
#2}}
\providecommand{\BIBdecl}{\relax}
\BIBdecl

\bibitem{mayer2012third}
J.~R. Mayer and J.~C. Mitchell, ``Third-party web tracking: Policy and
  technology,'' in \emph{IEEE S\&P}, 2012.

\bibitem{mikians2012detecting}
J.~Mikians, L.~Gyarmati, V.~Erramilli, and N.~Laoutaris, ``Detecting price and
  search discrimination on the internet,'' in \emph{HotNets}, 2012.

\bibitem{crowds:tissec}
M.~Reiter and A.~Rubin, ``Crowds: Anonymity for web transactions,'' \emph{ACM
  ToISS}, 1998.

\bibitem{tor-design}
R.~Dingledine, N.~Mathewson, and P.~Syverson, ``{Tor}: The second-generation
  onion router,'' in \emph{USENIX Security}, 2004.

\bibitem{tor-metrics}
{The Tor Project, Inc}, ``Tor metrics portal: Users,''
  \url{https://metrics.torproject.org/users.html}, accessed: 2014-02-11.

\bibitem{paul2011architectures}
S.~Paul, J.~Pan, and R.~Jain, ``Architectures for the future networks and the
  next generation internet: A survey,'' \emph{Computer Communications}, 2011.

\bibitem{find}
{The National Science Foundation}, ``{NSF} {NeTS} {FIND} initiative,''
  \url{http://www.nets-find.net/index.php}, accessed: 2014-02-11.

\bibitem{fire}
CORDIS, ``{FIRE} home page,''
  \url{http://cordis.europa.eu/fp7/ict/fire/home_en.html}, accessed:
  2014-02-11.

\bibitem{akari}
{National Institute of Information and Communications Technology}, ``{"AKARI"}
  architecture design project for new generation network,''
  \url{http://www.nict.go.jp/en/photonic_nw/archi/akari/akari-top_e.html},
  accessed: 2014-02-11.

\bibitem{papadopoulos2010greedy}
F.~Papadopoulos, D.~Krioukov, M.~Bogua, and A.~Vahdat, ``Greedy forwarding in
  dynamic scale-free networks embedded in hyperbolic metric spaces,'' in
  \emph{IEEE INFOCOM}, 2010.

\bibitem{bhattacharjee2006postmodern}
B.~Bhattacharjee, K.~Calvert, J.~Griffioen, N.~Spring, and J.~P. Sterbenz,
  ``Postmodern internetwork architecture,'' \emph{NSF Nets FIND Initiative},
  2006.

\bibitem{pathlet}
P.~B. Godfrey, I.~Ganichev, S.~Shenker, and I.~Stoica, ``Pathlet routing,'' in
  \emph{ACM SIGCOMM}, 2009.

\bibitem{rfc6830}
\BIBentryALTinterwordspacing
D.~Farinacci, D.~Lewis, D.~Meyer, and V.~Fuller, ``The locator/{ID} separation
  protocol ({LISP}),'' RFC 6830, 2013. [Online]. Available:
  \url{http://tools.ietf.org/html/rfc6830}
\BIBentrySTDinterwordspacing

\bibitem{yang2006source}
X.~Yang and D.~Wetherall, ``Source selectable path diversity via routing
  deflections,'' \emph{ACM SIGCOMM Computer Communication Review}, 2006.

\bibitem{yang2003nira}
X.~Yang, ``{NIRA}: A new internet routing architecture,'' in \emph{ACM SIGCOMM
  FDNA}, 2003.

\bibitem{zhang2011scion}
X.~Zhang, H.-C. Hsiao, G.~Hasker, H.~Chan, A.~Perrig, and D.~G. Andersen,
  ``{SCION}: Scalability, control, and isolation on next-generation networks,''
  in \emph{IEEE S\&P}, 2011.

\bibitem{geni}
A.~Falk, ``{GENI} at a glance,''
  \url{http://www.geni.net/wp-content/uploads/2011/06/GENI-at-a-Glance-1Jun201%
1.pdf}, 2011.

\bibitem{lap}
H.-C. Hsiao, T.-J. Kim, A.~Perrig, A.~Yamada, S.~C. Nelson, M.~Gruteser, and
  W.~Meng, ``{LAP}: Lightweight anonymity and privacy,'' in \emph{IEEE S\&P},
  2012.

\bibitem{anon_terminology}
A.~Pfitzmann and M.~Hansen, ``A terminology for talking about privacy by data
  minimization,''
  \url{http://dud.inf.tu-dresden.de/literatur/Anon_Terminology_v0.34.pdf},
  2010, v0.34.

\bibitem{ikev2}
\BIBentryALTinterwordspacing
C.~Kaufman, P.~Hoffman, Y.~Nir, and P.~Eronen, ``{Internet Key Exchange
  Protocol Version 2 (IKEv2)},'' RFC 5996 (Proposed Standard), Internet
  Engineering Task Force, Sep. 2010, updated by RFCs 5998, 6989. [Online].
  Available: \url{http://www.ietf.org/rfc/rfc5996.txt}
\BIBentrySTDinterwordspacing

\bibitem{eckersley2010unique}
P.~Eckersley, ``How unique is your web browser?'' in \emph{PETS}, 2010.

\bibitem{soltani2009flash}
A.~Soltani, S.~Canty, Q.~Mayo, L.~Thomas, and C.~J. Hoofnagle, ``Flash cookies
  and privacy,'' in \emph{SSRN eLibrary}, 2009.

\bibitem{econymics}
A.~Acquisti, R.~Dingledine, and P.~Syverson, ``On the economics of anonymity,''
  in \emph{FC}, 2003.

\bibitem{dingledine2009performance}
R.~Dingledine and S.~J. Murdoch, ``Performance improvements on {Tor} or, why
  {Tor} is slow and what we're going to do about it,''
  \url{http://www.torproject.org/press/presskit/2009-03-11-performance.pdf},
  2009.

\bibitem{jansenlira}
R.~Jansen, A.~Johnson, and P.~Syverson, ``{LIRA: Lightweight Incentivized
  Routing for Anonymity},'' in \emph{NDSS}, 2013.

\bibitem{dischinger2007characterizing}
M.~Dischinger, A.~Haeberlen, K.~P. Gummadi, and S.~Saroiu, ``Characterizing
  residential broadband networks,'' in \emph{ACM SIGCOMM IMC}, 2007.

\bibitem{levine2004timing}
B.~N. Levine, M.~K. Reiter, C.~Wang, and M.~Wright, ``Timing attacks in
  low-latency mix systems,'' in \emph{Proc. Financial Cryptography}, 2004, pp.
  251--265.

\bibitem{rainbow}
A.~Houmansadr, N.~Kiyavash, and N.~Borisov, ``{RAINBOW: A} robust and invisible
  non-blind watermark for network flows,'' in \emph{NDSS}, 2009.

\bibitem{voip06}
S.~Chen, X.~Wang, and S.~Jajodia, ``On the anonymity and traceability of
  peer-to-peer voip calls,'' \emph{Network, IEEE}, vol.~20, no.~5, pp. 32--37,
  2006.

\bibitem{sellingClickstream}
J.~Reimer, ``Your {ISP} may be selling your web clicks,''
  \url{http://arstechnica.com/tech-policy/2007/03/your-isp-may-be-selling-your%
-web-clicks/}, 2007.

\bibitem{cableOneSpy}
P.~Dampier, ```{Cable} {ONE} spied on customers' alleges federal class action
  lawsuit,''
  \url{http://stopthecap.com/2010/02/08/cable-one-spied-on-customers-alleges-f%
ederal-class-action-lawsuit}, 2012.

\bibitem{entropist}
P.~Syverson, ``Why {I'm} not an entropist,'' in \emph{Seventeenth International
  Workshop on Security Protocols.}\hskip 1em plus 0.5em minus 0.4em\relax
  Springer, 2009.

\bibitem{murdoch-pet2007}
S.~J. Murdoch and P.~Zieli{\'n}ski, ``Sampled traffic analysis by
  {I}nternet-exchange-level adversaries,'' in \emph{PETS}, 2007.

\bibitem{mti}
E.~MacAskill, J.~Borger, N.~Hopkins, N.~Davies, and J.~Ball, ``{GCHQ} taps
  fibre-optic cables for secret access to world's communications,''
  \url{http://www.theguardian.com/uk/2013/jun/21/gchq-cables-secret-world-comm%
unications-nsa}, June 2013.

\bibitem{undersea}
S.~Staff, ``Inside {TAO: Documents} reveal top {NSA} hacking unit,''
  \url{http://www.spiegel.de/international/world/the-nsa-uses-powerful-toolbox%
-in-effort-to-spy-on-global-networks-a-940969-3.html}, Dec. 2013.

\bibitem{rfc791}
\BIBentryALTinterwordspacing
J.~Postel, ``Internet protocol,'' RFC 791, 1981. [Online]. Available:
  \url{http://www.ietf.org/rfc/rfc0791.txt}
\BIBentrySTDinterwordspacing

\bibitem{bellovin1989security}
S.~M. Bellovin, ``Security problems in the {TCP/IP} protocol suite,'' \emph{ACM
  SIGCOMM CCR}, 1989.

\bibitem{rfc4271}
\BIBentryALTinterwordspacing
Y.~Rekhter, T.~Li, and S.~Hares, ``A border gateway protocol 4 ({BGP}-4),'' RFC
  4271, 2006. [Online]. Available: \url{http://tools.ietf.org/html/rfc4271}
\BIBentrySTDinterwordspacing

\bibitem{boyan1997anonymizer}
J.~Boyan, ``The anonymizer,'' \emph{Computer-Mediated Communication Magazine},
  1997.

\bibitem{freedom2-arch}
P.~Boucher, A.~Shostack, and I.~Goldberg, ``Freedom systems 2.0 architecture,''
  Zero Knowledge Systems, {Inc.}, White Paper, 2000.

\bibitem{PanchenkoPR08}
A.~Panchenko, L.~Pimenidis, and J.~Renner, ``Performance analysis of anonymous
  communication channels provided by {Tor},'' in \emph{ARES}, 2008.

\bibitem{andana}
S.~DiBenedetto, P.~Gasti, G.~Tsudik, and E.~Uzun, ``{ANDaNA: Anonymous} named
  data networking application,'' in \emph{NDSS}, 2013.

\bibitem{gao2001inferring}
L.~Gao, ``On inferring autonomous system relationships in the internet,''
  \emph{IEEE/ACM ToN}, 2001.

\bibitem{giotsas2012valley}
V.~Giotsas and S.~Zhou, ``Valley-free violation in internet routing-analysis
  based on {BGP} community data,'' in \emph{IEEE ICC}, 2012.

\bibitem{ixp}
P.~S. Ryan and J.~Gerson, ``A primer on {Internet} exchange points for
  policymakers and non-engineers,'' \url{http://ssrn.com/abstract=2128103},
  Aug. 2012.

\bibitem{lodhi2014open}
A.~Lodhi, A.~Dhamdhere, and C.~Dovrolis, ``Open peering by {Internet} transit
  providers: {Peer} preference or peer pressure?'' in \emph{Proc. IEEE
  INFOCOM}, 2014.

\bibitem{lamport1981password}
L.~Lamport, ``Password authentication with insecure communication,''
  \emph{Communications of the ACM}, 1981.

\bibitem{sdos}
N.~Borisov, G.~Danezis, P.~Mittal, and P.~Tabriz, ``Denial of service or denial
  of security?'' in \emph{CCS}, 2007.

\bibitem{wals08}
M.~K. Wright, M.~Adler, B.~N. Levine, and C.~Shields, ``Passive-logging attacks
  against anonymous communications systems,'' \emph{ACM Transactions on
  Information and System Security (TISSEC)}, vol.~11, no.~2, 2008.

\bibitem{ajaxchannels}
S.~Chen, R.~Wang, X.~Wang, and K.~Zhang, ``Side-channel leaks in web
  applications: {A} reality today, a challenge tomorrow,'' in \emph{IEEE S\&P},
  2010.

\bibitem{throughput}
P.~Mittal, A.~Khurshid, J.~Juen, M.~Caesar, and N.~Borisov, ``Stealthy traffic
  analysis of low-latency anonymous communication using throughput
  fingerprinting,'' in \emph{ACM CCS}, 2011.

\bibitem{latency}
N.~Hopper, E.~Y. Vasserman, and E.~Chan-Tin, ``How much anonymity does network
  latency leak?'' in \emph{ACM CCS}, 2007.

\bibitem{torta05}
S.~J. Murdoch and G.~Danezis, ``Low-cost traffic analysis of {Tor},'' in
  \emph{IEEE S\&P}, 2005.

\bibitem{congestion-longpaths}
N.~Evans, R.~Dingledine, and C.~Grothoff, ``A practical congestion attack on
  {T}or using long paths,'' in \emph{USENIX Security}, 2009.

\bibitem{serjantov2003entropy}
A.~Serjantov and G.~Danezis, ``Towards an information theoretic metric for
  anonymity,'' in \emph{PETS}, 2003.

\bibitem{caida}
{CAIDA}, ``The {CAIDA} {UCSD} inferred {AS} relationships - 20120601,''
  \url{http://www.caida.org/data/active/as-relationships/index.xml}, 2012.

\end{thebibliography}

\end{document}